\documentclass[final]{siamart250211}

\usepackage{amssymb}
\usepackage{hyperref}
\hypersetup{colorlinks,
	citecolor=blue,
	filecolor=black,
	linkcolor=black,
	urlcolor=blue
}

\usepackage{cleveref} % per the OP's example
\usepackage{float}
\usepackage{multirow}
\usepackage{tikz}
\usepackage{caption}
\usepackage{subfig}
\usepackage{algorithm}
\usepackage{algpseudocode}
\usepackage{listings}
\usepackage{xspace}
\usepackage{pgfplots}
\usepackage{multirow}
\usepackage{array}
\usepackage{empheq}
\usepackage{pgf-pie}
\usepackage{multirow}

\usepackage[figuresright]{rotating}
\usetikzlibrary{external}
\usepackage{soul}
\usepackage{todonotes}


\makeatletter
\renewcommand{\todo}[2][]{\tikzexternaldisable\@todo[#1]{#2}\tikzexternalenable}
\makeatother

\newcommand{\norm}[1]{\left\lVert#1\right\rVert}

\newcommand{\myint}{\int\limits}
\newcommand{\diff}[1]{\, d#1}
\newcommand{\vect}[1]{\boldsymbol{#1}}
\usepackage{mleftright}
\newcommand{\of}[1]{\mleft( #1 \mright)}

\newcommand{\reals}{\mathbb{R}}

\newcommand{\bte}{\textsc{Boltzsim}}
\newcommand{\vtheta}{{v_{\theta}}}
\newcommand{\vphi}{v_{\varphi}}

\newcommand{\bolsig}{\textsc{Bolsig+}\xspace}

\definecolor{codegreen}{rgb}{0,0.6,0}
\definecolor{codegray}{rgb}{0.5,0.5,0.5}
\definecolor{codepurple}{rgb}{0.58,0,0.82}
\definecolor{backcolour}{rgb}{0.95,0.95,0.92}

\lstdefinestyle{mystyle}{
  backgroundcolor=\color{backcolour}, commentstyle=\color{codegreen},
  keywordstyle=\color{magenta},
  numberstyle=\tiny\color{codegray},
  stringstyle=\color{codepurple},
  basicstyle=\ttfamily\footnotesize,
  breakatwhitespace=false,         
  breaklines=true,                 
  captionpos=b,                    
  keepspaces=true,                 
  numbers=left,                    
  numbersep=5pt,                  
  showspaces=false,                
  showstringspaces=false,
  showtabs=false,                  
  tabsize=2
}

\definecolor{a1}{RGB}{228,26,28}
\definecolor{a2}{RGB}{55,126,184}
\definecolor{a3}{RGB}{77,175,74}
\definecolor{a4}{RGB}{152,78,163}
\definecolor{a5}{RGB}{255,127,0}
\definecolor{a6}{RGB}{166,86,40}
\definecolor{a7}{RGB}{166,86,40}
\definecolor{a8}{RGB}{247,129,191}

\makeatletter
\DeclareRobustCommand{\pdot}{\mathbin{\mathpalette\pdot@\relax}}
\newcommand{\pdot@}[2]{%
  \ooalign{%
    $\m@th#1\circ$\cr
    \hidewidth$\m@th#1\cdot$\hidewidth\cr
  }%
}

\ifpdf
\DeclareGraphicsExtensions{.eps,.pdf,.png,.jpg}
\else
\DeclareGraphicsExtensions{.eps}
\fi

\headers{\tiny{\textsc{Boltzsim}: A fast solver for the 1D-space electron Boltzmann equation}}{\tiny{M. Fernando, J. Almgren-Bell, T. Oliver, R. Morser, P. Varghese, L. Raja, and G. Biros}}

\title{\textsc{Boltzsim}: A fast solver for the 1D-space electron Boltzmann equation with applications to radio-frequency glow discharge plasmas\thanks{Submitted to the editors February 23, 2025. \funding{This work was funded by the U.S. Department of Energy, National Nuclear Security Administration award number DE-NA0003969}}}

\author{ Milinda Fernando\thanks{Oden Institute, The University of Texas at Austin (\email{milinda@oden.utexas.edu}, \email{jalmgrenbell@utexas.edu}, \email{oliver@oden.utexas.edu}, \email{rmoser@oden.utexas.edu}, \email{varghese@mail.utexas.edu}, \email{lraja@mail.utexas.edu}, \email{biros@oden.utexas.edu}).}
		\and James Almgren-Bell\footnotemark[1]
		\and Todd Oliver\footnotemark[1] 
		\and Robert Moser\footnotemark[1]
		\and Philip Varghese\footnotemark[1]
		\and Laxminarayan Raja\footnotemark[1]
		\and George Biros\footnotemark[1]}

\ifpdf
\hypersetup{
  pdftitle={\textsc{Boltzsim}: A fast solver for the 1D-space electron Boltzmann equation with applications to radio-frequency glow discharge plasmas},
  pdfauthor={M. Fernando, J. Almgren-Bell, T. Oliver, R. Morser, P. Varghese, L. Raja, and G. Biros}
}
\fi

\begin{document}
\maketitle

% REQUIRED
\begin{abstract}
We present an algorithm for solving the one-dimensional space collisional Boltzmann transport equation (BTE) for electrons in low-temperature plasmas (LTPs). Modeling LTPs is useful in many applications, including advanced manufacturing, material processing, and hypersonic flows, to name a few. The proposed BTE solver is based on an Eulerian formulation. It uses Chebyshev collocation method in physical space and a combination of Galerkin and discrete ordinates in velocity space. We present self-convergence results and cross-code verification studies compared to an in-house particle-in-cell (PIC) direct simulation Monte Carlo (DSMC) code. \bte~is our open source implementation of the solver. Furthermore, we use \bte~to simulate radio-frequency glow discharge plasmas (RF-GDPs) and compare with an existing methodology that approximates the electron BTE. We compare these two approaches and quantify their differences as a function of the discharge pressure. The two approaches show an 80x, 3x, 1.6x, and 0.98x difference between cycle-averaged time periodic electron number density profiles at 0.1 Torr, 0.5 Torr, 1 Torr, and 2 Torr discharge pressures, respectively. As expected, these differences are significant at low pressures, for example less than 1 Torr. 

%the electron BTE with the widely used two temperature ``fluid'' approximations with tabulated coefficients. We report the differences between the electron BTE and fluid approximation of electron transport. As expected, these differences are significant in lower pressures (i.e., less than 200 mTorr). %especially in low-pressure RF-GDPs. 

%Most state-of-the-art methods for electron kinetics are based on Monte-Carlo sampling for collisions combined with Lagrangian particle-in-cell methods. We discuss
%an Eulerian solver that approximates the electron velocity distribution function using spherical harmonics (angular components)
%and B-splines (energy component). Our solver supports electron-heavy elastic and inelastic binary collisions, electron-electron
%Coulomb interactions, steady-state and transient dynamics, and an arbitrary nmber of angular terms in the electron distribution
%function. We report convergence results and compare our solver to two other codes: an in-house particle Monte-Carlo method; and
%Bolsig+, a state-of-the-art Eulerian solver for electron transport in LTPs. Furthermore, we use our solver to study the relaxation
%time scales of the higher-order anisotropic correction terms. Our code is open-source and provides an interface that allows coupling
%to multiphysics simulations of low-temperature plasmas.
\end{abstract}

% REQUIRED

% REQUIRED

\section{Introduction} \label{sec:intro}
%The plasma state can be considered as the fourth state of matter, and it is characterized by the collective behavior of its species (i.e., neutrals, ions, and electrons). Predictive mathematical modeling of plasmas is useful in many application domains in science and engineering.
Low-temperature plasmas (LTPs) find application in advan-ced manufacturing, semiconductor processing, the medical device industry, plasma-assisted combustion, and numerous other application domains. Modeling of low-temperature, weakly ionized plasmas can be challenging due to non-equilibrium chemistry, time-scale differences, and non-local electromagnetic coupling of charged particles. Electrons in LTPs often deviate from thermal equilibrium. Hence, their velocity space distribution functions differ from the Maxwell-Boltzmann distribution. Accurate modeling of LTPs requires solving the BTE for the electron distribution function (EDF). The BTE evolves the EDF dynamics and is coupled to ion and neutral species dynamics. Solving the BTE is a challenging task due to the dimensionality of the problem, the multiple time scales in various plasma processes, and dense collisional operators. Predictive simulation of LTPs relies on a hierarchy of models ranging from semi-analytical models to a fully kinetic treatment of plasma. Most existing LTP simulations rely on a model known as the ``fluid approximation'' that is derived from averaging the first three velocity space moments, mass, momentum, and energy of the BTE. The fluid approximation for the electron transport is based on assumptions on the EDF that may be invalid in certain LTP conditions. They also require approximating quantities that depend on the EDF, e.g., reaction rates, electron mobility, and electron diffusivity.
%, which is globally coupled with self-consistent electromagnetic effects. Representing an arbitrary EDF necessitates solving the electron BTE for specified LTP conditions.  

In this paper, we relax the fluid approximation for the electrons and discuss the numerical solution of the BTE in one spatial dimension and three velocity dimensions. We refer to this problem as the 1D3V BTE and the spatially homogeneous case as the 0D3V BTE. There are two main approaches for numerically solving the 1D3V BTE: Lagrangian and Eulerian. In the Lagrangian approach, a particle-in-cell (PIC) scheme is used for phase space advection, and the direct simulation Monte Carlo (DSMC) sampling approach is used to model particle collisions. In contrast, Eulerian solvers discretize the BTE in a fixed frame with standard PDE discretization techniques such as Galerkin, finite differences, and spectral methods. Eulerian BTE solvers are also referred to as deterministic BTE solvers because of the absence of sampling methods for particle collisions. In this paper, we focus on the Eulerian approach for solving spatially coupled electron BTE for LTPs. The developed methodology is applied to the kinetic treatment of electrons in low-temperature radio-frequency glow discharge plasmas (RF-GDPs). To our knowledge, existing work uses either PIC-DSMC or fluid approximations for RF-GDP simulations. We introduce a hybrid solver where the heavies are represented using a fluid approximation, and the electrons are by the BTE. We refer to the new solver as~\bte. Our contributions are summarized below. 
%GDPs have applications in plasma cleaning, material processing, and plasma chemistry. 
%The current existing modeling methodology for RF-GDPs relies on fluid models with drift-diffusion approximation for species transport. The developed BTE solver is used to create a hybrid model for RF-GDPs. In the proposed model, electrons are modeled using the BTE, and heavy species (i.e., ions) are modeled using the standard fluid approximation. The contributions of the presented work are summarized below.  

\begin{itemize}
	\item \textbf{Multi-term EDF approximation}: \bte~extends the traditional two-term EDF expansion with arbitrary multi-term EDF approximations for one-dimensional space electron BTE. 
	\item \textbf{Semi-implicit and fully implicit schemes}: We present novel numerical schemes for the time evolution of electron BTE, with semi-implicit and fully-implicit coupling with the self-consistent electric field generated by charged plasma species. We propose a domain decomposition-based preconditioning method for the implicit time integration of the BTE  for both semi and fully implicit schemes.  
	\item \textbf{Self-consistent hybrid model for RF-GDPs}: We propose a self-con-sistent fluid and BTE combined hybrid transport model for RF-GDPs. We use the BTE for the kinetic treatment of electrons and the fluid approximation for heavy species transport. 
	\item \textbf{Comparison between fluid and hybrid models}: A comparison between fluid and hybrid transport models is carried out for RF-GDPs, with argon ions and electrons. Using the RF-GDP setup, we verify our solver by comparing to an in-house PIC-DSMC code. 
	\item \textbf{Open source}: Our solver is implemented in the Python programming language with support for CPU and portable GPU acceleration. The solver is openly available at \url{https://github.com/ut-padas/boltzmann}. 
\end{itemize}
%In our experiments, we use a chemistry model in which the neutral number density is constant and with electron-neutral momentum transfer and ionization reactions. So, the heavies consist of ions. Our methodology, however, readily extends to more general chemistry models.  
%
\section{Background}
\label{sec:background}
%There exist several Eulerian codes~\cite{hagelaar2005solving,hagelaar2015coulomb,frost1962rotational,COMSOL1998, pancheshnyi2008zdplaskin,morgan1990elendif} that use fixed two-term EDF approximations for solving spatially homogeneous BTE. Similar to~\cite{fernando0DBTE} the work in \cite{mutibolt} also supports multi-term EDF approximations. However, it does not support Coulomb interactions and transient solutions. The PIC-DSMC approach is widely used for solving spatially inhomogeneous BTE for various application settings~\cite{scanlon2010open, zabelok2015adaptive, frezzotti2011solving, oblapenko2020velocity, bartel2003modelling, lymberopoulos1994stochastic, levko2021vizgrain}. To the best of our knowledge, there has been limited work on spatially inhomogeneous Eulerian electron BTE solvers for LTPs as well as RF-GDPs. Also, the existing literature has limited discussion on solver algorithmic details and related complexity analysis for 1D3V LTP solvers.

We summarize existing methods for modeling GDPs. The fluid approximation is the most widely used approach to simulate RF-GDPs~\cite{boeuf1987numerical,barnes1988staggered,zhao2018numerical,young1993two, young1993comparative,panneer2015computational,kothnur2007simulation,liu2014numerical}. 
%The fluid approximation for species transport is derived from the BTE's mass, momentum, and energy moments evolution equations. The momentum continuity equation is further simplified using the so-called drift-diffusion~\cite{hill1986introduction} approximation. The drift-diffusion approximation eliminates the time evolution of the momentum continuity equation, which is indirectly coupled to the mass continuity equation. The work described by \cite{zhao2018numerical, liu2014numerical} presents a detailed study on the effect of the secondary electron emission coefficient using the one-dimensional fluid approximation of RF-GDPs. 
There have been several attempts for the two-dimensional modeling of RF-GDPs using the fluid approximation~\cite{young1993two, wilcoxson1996simulation}. 
%The work described in \cite{lin2001simulation} presents a reduced-order modeling approach for RF-GDPs, based on the solutions computed by the fluid approximation. 
Lagrangian particle-in-cell algorithms for evolving fluid equations for GDPs are discussed in~\cite{young1993two, young1993comparative}. The fluid approximation assumes that species distribution functions are isotropic. This is not the case, especially for the electrons closer to discharge walls due to a combination of boundary conditions and the strong electric field gradients in the sheath. Most BTE solvers are based on the PIC-DSMC approach~\cite{scanlon2010open, zabelok2015adaptive, frezzotti2011solving, oblapenko2020velocity, bartel2003modelling, lymberopoulos1994stochastic, levko2021vizgrain}. 
There have been several attempts~\cite{surendra1991particle, bogaerts1999hybrid, satake1997two, tsendin1995electron, yuan20171d} to develop more accurate models for GDPs, specifically using the BTE for electrons.
The authors in~\cite{surendra1991particle} use a PIC-DSMC approach to model species transport using the BTE but have limited details on the collisional process. A hybrid modeling approach that uses the fluid approximation for heavies and the BTE for electrons is proposed in~\cite{bogaerts1999hybrid} using a PIC-DSMC to simulate the BTE effects.
%In the above, their PIC-DSMC implementation requires an electric field as a spatiotemporal field, which is computed from the fluid approximation. 
In~\cite{satake1997two} the authors propose a two-dimensional hybrid model for RF-GDPs by evolving the BTE with a simplified Bhatnagar–Gross–Krook (BGK) collision operator. 
A hybrid transport model with two-term EDF approximations is proposed in~\cite{ yuan20171d,loffhagen2009advances}. The work given by~\cite{yuan20171d} uses an approximation scheme to update the two EDF components, and the work by~\cite{loffhagen2009advances} uses a quasi-stationary approximation of the EDF with a simplified collision operator. Both of these works do not solve the one-dimensional electron BTE.

To the best of our knowledge, there has been limited work on spatially inhomogeneous Eulerian electron BTE solvers for LTPs as well as RF-GDPs. Also, the existing literature has limited discussion on solver algorithmic details and related complexity analysis for 1D3V LTP solvers. Our work is based on a Eulerian 0D3V BTE solver we developed in our group~\cite{fernando0DBTE}. For related work please see the discussion in~\cite{fernando0DBTE}. Here we focus on 1D3V solvers.
\section{Methodology} \label{sec:methodology}

%In low-temperature plasmas (LTPs), electrons often deviate from the Maxwell-Boltzmann (Maxwellian) distribution. An accurate representation of the electron distribution function (EDF) is critical for accurate LTP modeling. Representing such non-Maxwellian distribution functions requires one to solve the Boltzmann transport equation (BTE). 

This section presents an overview of the electron BTE (see \Cref{subsec:bte}) and its application to model electron transport in glow discharge devices (see \Cref{subsec:glow_discharge}). 
\Cref{tab:summary_symbols} summarizes the symbols used in the paper. 
%presents symbols used in this paper and their descriptions.
\begin{table}[!tbhp]
	\centering
	\resizebox{0.9\textwidth}{!}{
		\begin{tabular}{||c|p{6cm}|c|p{7cm}||}
			\hline
			\hline
			Symbol & Description 	&		Symbol & Description\\
			\hline
			$p_0$    & \small Gas pressure 	  & $T_0$    & \small Gas temperature\\
			$n_0$    & \small Neutral density & $L  $    & \small Electrode gap \\
			$V_0$    & \small Peak voltage    & $\zeta  $    & \small Oscillation frequency\\
			$D_e$    & \small Electron diffusion coefficient & $D_i$    & \small Ion diffusion coefficient \\
			$\mu_e$  & \small Electron mobility coefficient & $\mu_i$  & \small Ion mobility coefficient \\
			$T_e$    & \small Electron temperature & $q_e$    & \small Electron charge\\
			$m_e$    & \small Electron mass      & $f\of{\vect{x}, \vect{v}, t}$ & \small EDF \\
			$\vect{E}$ & \small Electric field   & $N_x$ & \small Number of collocation points in $x$\\ 
			$N_r$ & \small Number of B-splines in v & $N_\vtheta$ & \small Number of ordinates in $\vtheta$\\
			$N_l$ & \small Number of spherical harmonics in $\vtheta$ & $\vect{C}_{en}$ & \small Discrete electron-neutral collision operator \\ 
			$\vect{A}_v$ & \small Discrete velocity space advection & $\vect{D}_x$ & \small Spatial derivative operator (Chebyshev-collocation)\\
			$\vect{A}_x$ & \small Galerkin projection of $v \cos(\vtheta) \nabla_{\vect{x}} f$ term & $\vect{P}_S$ & \small $\vtheta$ ordinates to spherical harmonics projection \\
			$\vect{P}_O$ & \small Spherical harmonics to $\vtheta$ ordinates projection & $\vect{F}$ & \small Discretized EDF \\
			$\sigma_{0}$ & \small Momentum transfer cross-section & $\sigma_{i}$ & \small Ionization cross-section \\
			\hline
			\hline
	\end{tabular}}
	\caption{Notation \label{tab:summary_symbols}}
\end{table}
\subsection{The electron Boltzmann transport equation (BTE)}
\label{subsec:bte}
The electron BTE driven by an electric field $\vect{E}\of{\vect{x}, t}$ is given by
\begin{equation}\label{eq:bte-full}
	\partial_t f + \vect{v} \cdot \nabla_{\vect{x}} f- \frac{q_e \vect{E}}{m_e} \cdot \nabla_{\vect{v}} f = C_{en}(f).
\end{equation} Here, $f\of{\vect{x}, \vect{v}, t}$ denotes the EDF, $\vect{x} \in \Omega \subset \reals^3$, $\vect{v}\in \reals^3$ and $t\in \reals$, and $C_{en}$ denotes the electron-heavy  binary collision operator. %The evolution of the EDF is governed by the integro-differential Boltzmann equation given by \Cref{eq:bte-full}. In \Cref{eq:bte-full} the left-hand side (LHS) denotes the spatial advection (i.e., $\vect{v} \cdot \nabla_{\vect{x}} f$) velocity space acceleration (i.e., $\frac{q_e \vect{E}}{m_e} \cdot \nabla_{\vect{v}} f$),  and the right-hand side denotes the electron-heavy binary collisions, e.g., elastic, ionization. 
$C_{en}$ compromises the electron-heavy momentum transfer and ionization collisions and $f\of{\vect{x}, \vect{v}, t} d\vect{x} d\vect{v}$ denotes the number of electrons for the phase-space volume $d\vect{x} d\vect{v}$. Using spherical coordinates $\vect{v}=\of{v, \vtheta, \vphi}$ in velocity space and aligning the velocity space z-axis to $\vect{E}$ direction, we obtain $\vect{E} = E \vect{\hat{e}_z} = E \of{\cos(\vtheta) \vect{\hat{e}_r} - \sin(\vtheta)\vect{\hat{e}_\theta}}$. By restricting $\vect{x}$ in 1D, we obtain the 1D3V BTE which is given by
% The 1D BTE is given by 
%we can write the 1D-space Boltzmann equation as in \Cref{eq:bte-1d}.
\begin{equation}
    \text{[1D3V]:} \quad \partial_t f + v\cos\of{\vtheta} \partial_x f- \frac{q_e E}{m_e} \of{\cos(\vtheta) \partial_v f - \sin(\vtheta) \frac{1}{v} \partial_{\vtheta}f } = C_{en}(f). \label{eq:bte-1d}
\end{equation}
Furthermore, if we drop the spatial dependence entirely, we obtain the homogeneous 0D3V case, which is given by
\begin{equation}
	\text{[0D3V]:} \quad \partial_t f - \frac{q_e E}{m_e} \of{\cos(\vtheta) \partial_v f - \sin(\vtheta) \frac{1}{v} \partial_{\vtheta}f } = C_{en}(f). \label{eq:bte-0d}
\end{equation}

\par \textbf{Electron-heavy collisions}: The electron-heavy collision operator $C_{en}$ captures the rate of change in the EDF due to collisions. It is given by 
%\begin{multline}
%	C^{-}_{en}(f,f_0) = 
%	\qquad\myint_{\reals^3} \myint_{\reals^3} \myint_{S^2} 
%	B\of{\vect{v}_e^\prime, \vect{v}_0^\prime, \vect{\omega}} 
%	f\of{\vect{v}_e^\prime} f_0\of{\vect{v}_0^\prime} 
%	\delta\of{\vect{v}_e^\prime - \vect{v}_e} 
%	\diff{\vect{v}_0^\prime} \diff{\vect{v}_e^\prime} \diff{\vect{\omega}}, \label{eq:c_en_loss}
%\end{multline}
\begin{align}
	&C^{-}_{en}(f,f_0) =\myint_{\reals^3} \myint_{\reals^3} \myint_{S^2} 
	B\of{\vect{v}_e^\prime, \vect{v}_0^\prime, \vect{\omega}} 
	f\of{\vect{v}_e^\prime} f_0\of{\vect{v}_0^\prime} 
	\delta\of{\vect{v}_e^\prime - \vect{v}_e} 
	\diff{\vect{v}_0^\prime} \diff{\vect{v}_e^\prime} \diff{\vect{\omega}}, \label{eq:c_en_loss}
	\\
	&C^{+}_{en}\of{f,f_0} =\myint_{\reals^3} \myint_{\reals^3} \myint_{S^2} 
	B\of{\vect{v}_e^\prime, \vect{v}_0^\prime, \vect{\omega}} 
	f\of{\vect{v}_e^\prime} f_0\of{\vect{v}_0^\prime} 
	\delta\of{\vect{v}_e^\text{post}\of{\vect{v}_e^\prime, \vect{v}_0^\prime, \vect{\omega}} - \vect{v}_e} 
	\diff{\vect{v}_0^\prime} \diff{\vect{v}_e^\prime} \diff{\vect{\omega}} \label{eq:c_en_gain},\\
	&C_{en}\of{f,f_0} = \qquad C^{+}_{en}\of{f,f_0} - C^{-}_{en}\of{f,f_0} \label{eq:c_en}.
\end{align} Here, $f_0$ denotes the heavy species distribution function, and $B$ denotes the probability measure for a given collision parameters. The collision kernel $B$ is defined by the collision cross-section~\cite{pitchford2017lxcat}. These expressions can be further simplified by using the following common to LTPs assumptions: 1) The heavy species distribution functions are Maxwellian, and 2) electron-heavy collisions follow isotropic scattering. Here, the isotropic scattering assumption ensures that the collision probability measure is independent of the scattering angle. Under these assumptions, the BTE preserves the azimuthal symmetry in the velocity space. \Cref{eq:c_en} can be further simplified by assuming Maxwellian distribution for heavy species, i.e., $f_{0}\of{\vect{v}} = \frac{n_0}{\of{\sqrt \pi v_{th,0}}^3} \exp\of{\of{-v/v_{th,0}}^2}$ where $v_{th,0}=\sqrt{2k_B T_0/m_0}$ is the thermal velocity, $n_0$ is the number density, and $m_0$ is the mass of the heavy species.      

% \textbf{Electron-heavy collisions}: In \Cref{eq:bte-1d}, $C_{en}$ denotes the integral operator capturing the binary collisions between electrons and background gas particles. 
\subsection{Glow discharge}
\label{subsec:glow_discharge}
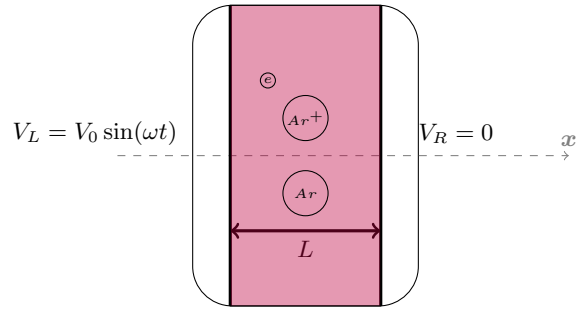
\begin{figure}[!tbhp]
    \centering
    \begin{tikzpicture}
        \draw[rounded corners=15pt]   (0,0) rectangle ++(3,4);
        \draw[->, dashed, gray]   (-1,2) -- (5,2) node[above] {\small $\vect{x}$};
        \draw[-,black, very thick]   (0.5,0)--(0.5,4) node at (-1.3, 2.3) {\small $V_L=V_0\sin\of{\omega t}$};
        \draw[-,black, very thick]   (2.5,0)--(2.5,4) node at ( 3.5, 2.3) {\small $V_R=0$};
        \draw[<->,black, very thick]   (0.5,1)  --(2.5,1) node[below, xshift=-1cm] {\small $L$};
        \draw[fill=purple, fill opacity=0.4]   (0.5,0) rectangle ++(2.0,4);
        \draw (1.5,1.5) circle [radius=0.3] node {\tiny $Ar$};
        \draw (1.5,2.5) circle [radius=0.3] node {\tiny $Ar^{+}$};
        \draw (1.0,3.0) circle [radius=0.1] node {\tiny $e$};
        %\node (1) [draw, rounded rectangle] {rounded rectangle};
    \end{tikzpicture}
    \caption{An illustrative schematic for the glow discharge plasma with Argon. \label{fig:glow_schematic}}
\end{figure}

Here, we summarize the formulation for RF-GDPs. An externally imposed potential difference between two electrodes at the opposite ends of a tube filled with argon is used to generate a plasma. \Cref{fig:glow_schematic} shows a schematic of the glow discharge device. 
The imposed potential generates an electric field that causes electron acceleration and ionization of the background heavy species, which results in the glow discharge plasma. Following standard approximations for GDPs, our formulation is based on the following assumptions:
%The following assumptions are made for the mathematical models used in this work to model the glow discharge phenomenon. 
\begin{enumerate}
    \item We assume the size of the electrodes is larger than the discharge length ($L$). Hence, discharge characteristics vary along the normal direction to electrodes.
    \item We limit our study to the three-species collisional model described in \Cref{t:col_list}. We only consider direct ionization from the neutral state, ignoring excitation and step-ionization from metastable states. We remark that our solver easily extends to more species and collision types. Furthermore, assuming constant $n_0(x, t)=n_0$ value, we only track ions and electrons. 
    \item We assume heavies have constant temperature $T_0$, typically the room temperature.  %\todo{do we need to mention on constant neutral density ? }
    \item Following~\cite{liu2014numerical}, we assume zero secondary electron emission from the electrode surface. 
\end{enumerate}
\begin{table}[!tbhp]
	\centering
	\begin{tabular}{|c|c|c|}
		\hline 
		Collision & Threshold & Post-collision\\
		\hline 
		$e + Ar \rightarrow e + Ar$ & None & $\varepsilon_{\text{post}} =\varepsilon_{\text{pre}}(1- \frac{2m_e}{m_0})$\\
		\hline
		$e + Ar \rightarrow Ar^{+} + 2e$ & $\varepsilon_{ion}$=15.76eV &  $\varepsilon_{\text{post}}=0.5 \of{\varepsilon_{\text{pre}}-\varepsilon_{\text{ion}}}$\\     
		\hline
	\end{tabular}
	\caption{Three species Ar collisional model. \label{t:col_list}}
\end{table}
We consider two formulations: In~\Cref{subsubsec:fluid_only}, we give a species transport formulation for the fluid approximation for both electrons and ions. In~\Cref{subsubsec:hybrid}, we give the hybrid approximation where the fluid approximation describes ions, and the BTE models the electron transport. As mentioned, we assume constant neutral species density and temperature in both cases.    
\subsubsection{Fluid model for electrons and ions}
\label{subsubsec:fluid_only}
We use a standard self-consistent fluid model for glow discharge phenomena~\cite{panneer2015computational,liu2014numerical}. The species number density continuity equations are given by 
\begin{subequations}
	\begin{align}
		&\partial_t{n_{e}} + \nabla \cdot \vect{J_e} = k_i n_0 n_e, \label{eq:fl_mass_continuity_ne}\\
		&\partial_t{n_{i}} + \nabla \cdot \vect{J_i} = k_i n_0 n_e. \label{eq:fl_mass_continuity_ni}
	\end{align}\label{eq:fl_mass_continuity}
\end{subequations}
\Cref{eq:fl_mass_continuity_ne,eq:fl_mass_continuity_ni} describe the evolution of ions and electron number densities where $k_i$ denotes the ionization rate coefficient, $n_0$ denotes the background neutral number density, and $n_e$ denotes the electron number density.  
%\begin{equation}
%    \partial_t{n_{k}} + \nabla \cdot \vect{J_k} = \dot{S}_k. \label{eq:fl_mass_continuity}
%\end{equation}
In \Cref{eq:fl_mass_continuity}, flux terms, $\vect{J}_{e}$ and $\vect{J}_{i}$ are approximated using a drift-diffusion approximation~\cite{hill1986introduction} and are given by 
\begin{equation}
	\vect{J}_e = -\mu_e n_e \vect{E} -D_e \nabla n_e \text{  and  } \vect{J}_i = \mu_i n_i \vect{E} -D_i \nabla n_i \label{eq:drift_diffusion_flux}.
\end{equation}
%In \Cref{eq:fl_mass_continuity_ne,eq:fl_mass_continuity_ni}, $\dot S_{e}$ and $\dot S_{i}$ denote the rate of production for electrons and ions. The above can be written as $\dot{S}_{e, i} = k_i n_0 n_e$, where $k_i$ denotes the ionization rate coefficient, $n_0$ denotes the background neutral number density, and $n_e$ denotes the electron number density.  
 
%Electrons undergo significant acceleration due to the electric field. The evolution of electron energy is prescribed by
The electron energy equation is given by 
\begin{equation}
	\partial_t \of{\frac{3}{2} n_e k T_e} = -\nabla \cdot \vect{q}_e - e \vect{J}_e \cdot \vect{E} - \varepsilon_{\text{ion}} k_i n_0 n_e \label{eq:fl_electron_energy} ,
\end{equation} and the electron energy flux term is specified by
\begin{equation}
	\vect{q_e} 	= -\frac{3 k D_e n_e}{2} \nabla T_e + \frac{5}{2} k T_e \vect{J_e} \label{eq:fl_e_energy_flux}.
\end{equation}

The underlying electric field potential is given by Gauss's law and given by
%specified by \Cref{eq:gauss_law}.
\begin{equation}
    \Delta V = -\frac{e}{\epsilon_0} (n_i-n_e) \text{ and } \vect{E} = - \nabla V \label{eq:gauss_law}.
\end{equation}
%%%%%%%%%%%%%%%%%%%%%%%%%%%%%%%%%%%%%%%%%%%%%%%%%%%%%%%%%%%%%%%%%%
\par \textbf{Boundary conditions}: For electrons, we enforce the Maxwellian flux boundary condition~\cite{panneer2015computational} given by
\begin{equation}
	\vect{J_e} \cdot \hat{\vect{n}} = \frac{1}{4} n_e \of{\frac{8k_B T_e}{\pi m_e}}^{0.5} \label{eq:e_flux_bc}, 
\end{equation} where $\hat{\vect{n}}$ is the outward unit normal vector. For the electron energy equation, the electron energy flux $\vect{q_e}$ at the boundary is enforced to the average energy each electron carry when it escapes from the discharge walls and given by $\vect{q_e} = \frac{5}{2} k_B T_e \vect{J_e}$. For the ions, the enforced boundary condition is given by
\begin{equation}
    \vect{J_i}\cdot \hat{\vect{n}} = - n_i \max(0, \mu_i \vect{E} \cdot \hat{\vect{n}}) \label{eq:fl_ion_bc}.
\end{equation} 
A Dirichlet condition for \Cref{eq:gauss_law} is given by the driving oscillatory voltage input
\begin{equation}
    V\of{x=0, t}=V_0 \sin\of{2\pi \zeta t} \text{ and } V\of{x=L, t} = 0 \label{eq:voltage_bc}.
\end{equation} 

The fluid approximation for the glow discharge problem requires the kinetic coefficients, $\mu_{e,i}$, $D_{e,i}$ and $k_i$. Following~\cite{liu2014numerical}, we use constant kinetic coefficients for the ions. For the electrons, we use temperature-based tabulated kinetic coefficients. \Cref{subsec:results_glowd} presents a detailed description of temperature-based electron kinetic coefficients. The fluid approximation for GDP simulation is summarized below. 
%\Cref{eq:fluid_model_eqs} summarizes the fluid model for glow discharge phenomena with the boundary conditions described above. 

\begin{subequations}
	\label{eq:fluid_model_eqs}
	\begin{empheq}[left={\forall t>0, x \in [0, L]\empheqlbrace}]{align}
			\partial_t n_i &= k_i(T_e) n_0 n_e - \nabla \cdot \vect{J_i} \label{e:fl_a},\\
			\partial_t n_e &= k_i(T_e) n_0 n_e - \nabla \cdot \vect{J_e} \label{e:fl_b},\\
			\partial_t \of{\frac{3}{2} n_e k T_e} &= -\nabla \cdot \vect{q}_e - e \vect{J_e} \cdot \vect{E} - \varepsilon_{\text{ion}} k_i(T_e) n_0 n_e \label{e:fl_c}, \\
			\vect{J}_i &=  \mu_i n_i \vect{E} -D_i \nabla n_i \label{e:fl_a_flux},\\
			\vect{J}_e &= -\mu_e\of{T_e} n_e \vect{E} -D_e\of{T_e} \nabla n_e\label{e:fl_b_flux},\\
			\vect{q_e} &= -\frac{3 k D_e\of{T_e} n_e}{2} \nabla T_e + \frac{5}{2} k T_e \vect{J_e}\label{e:fl_c_flux},\\
			\Delta V &= -\frac{e}{\epsilon_0} (n_i-n_e) \text{ , } \vect{E} = - \nabla V \label{e:fl_d}, 
	\end{empheq}
	\begin{empheq}[left={\forall t>0, x=0 \text{ and } x=L \empheqlbrace}]{align}
		& \vect{J_i}^\cdot \hat{\vect{n}} = - n_i \max(0, \mu_i \vect{E} \cdot \hat{\vect{n}}),\\
		& \vect{J_e} \cdot \hat{\vect{n}} = \frac{1}{4} n_e \of{\frac{8k_B T_e}{\pi m_e}}^{1/2} \label{e:fl_b_bdy},\\
		& \vect{q_e} = \frac{5}{2} k_B T_e \vect{J_e} \label{e:fl_c_bdy},\\
		&V\of{x=0, t}=V_0 \sin\of{2\pi \zeta t} \text{ , } V\of{x=L, t} = 0.
	\end{empheq}		
\end{subequations}

\subsubsection{Summary of the hybrid model}
\label{subsubsec:hybrid}
We replace Equations \eqref{e:fl_b}, \eqref{e:fl_c}, \eqref{e:fl_b_bdy}, and \eqref{e:fl_c_bdy}  with the 1D3V BTE. The combined equations become
%In the hybrid formulation, we use the fluid approximation for heavies and the BTE for electron transport. The overall hybrid model for the one-dimensional glow discharge problem is summarized by  
\begin{subequations}
	\label{eq:glow_hybrid}
	\begin{empheq}[left={\forall t>0, x \in [0, L]\empheqlbrace}]{align}
			& \partial_t{n_{i}} + \nabla \cdot \vect{J_{i}} = k_i n_0 n_e \label{eq:hybrid_cnt_a},\\
			&\partial_t f + v\cos\of{\vtheta} \partial_x f- \nonumber \\
			&\qquad\qquad\frac{q_e E}{m_e} \of{\cos(\vtheta) \partial_v f - \sin(\vtheta) \frac{1}{v} \partial_{\vtheta}f } = C_{en}(f) \label{eq:hybrid_cnt_b},  \\
			&\vect{J}_i =  \mu_i n_i \vect{E} -D_i \nabla n_i \label{eq:hybrid_cnt_a_flux},\\
			&k_i = \int_{\vect{v}} \norm{\vect{v}} \sigma_i \hat{f} \diff{\vect{v}} \text{ where } \hat{f}\of{\vect{v}} = \frac{f\of{\vect{v}}}{\int_{\vect{v}} f\of{\vect{v}} \diff{\vect{v}}}, \\
			& \Delta V = -\frac{e}{\epsilon_0} (n_i- n_e) \text{ , } \vect{E} = - \nabla V \label{eq:hybrid_cnt_c}, 
	\end{empheq} 
	\begin{empheq}[left={\forall t>0, x=0 \text{ and } x=L \empheqlbrace}]{align}
		& \vect{J_i}^\cdot \hat{\vect{n}} = - n_i \max(0, \mu_i \vect{E} \cdot \hat{\vect{n}}), \\
		& f(x=0, v, \vtheta \leq \frac{\pi}{2}, \vphi) = 0 \text{ and } \nonumber\\
		& \qquad\qquad f(x=L, v, \vtheta > \frac{\pi}{2}, \vphi) = 0 \label{eq:hybrid_cnt_b_bdy},\\
		& V\of{x=0, t}=V_0 \sin\of{2\pi \zeta t} \text{ , } V\of{x=L, t} = 0.
	\end{empheq}		
\end{subequations} Note that the EDF carries information on the electron number density $n_e(t,\vect{x}) = \int_{\vect{v}} f(t, \vect{x}, \vect{v}) \diff{\vect{v}}$ and $\sigma_i$ denotes the total ionization cross section and $\hat{f}$ denotes the normalized EDF. As prescribed above, we assume zero incoming flux boundary conditions for \Cref{eq:hybrid_cnt_b}.

\subsection{Numerical solver for the fluid approximation}
\label{subsec:fluid_solver}
As our focus is on the numerics of the hybrid solver, here we give a high-level description of the fluid solver. We are not claiming that this is the most efficient solver for this problem. Instead, we focus on comparing the converged time-periodic solutions between the fluid and the hybrid approaches. For the results of the fluid solver, we performed self-convergence tests in space and time. Now let us describe the solver for the fluid model described in \Cref{eq:fluid_model_eqs}. We use a Chebyshev collocation method for the spatial discretization. These collocation points are clustered to the electrodes. This is ideal for efficiently resolving sharp gradients of the electric field, species densities, and the electron energy in the sheath.  
%sheath profiles with sharp gradients near the electrodes. 
For the time discretization, we found that the operator-split approximation-based semi-implicit methods for solving~\Cref{e:fl_b,e:fl_c,e:fl_d} have timestep size restrictions determined by the electron timescales (see \Cref{subsubsec:ts_analysis}). Therefore, the system given in \Cref{eq:fluid_model_eqs} is solved as a coupled non-linear system with fully implicit time integration. We use Newton's method with direct factorization of the assembled Jacobian to solve for the Newton step. 
%The overall computational complexity for a single timestep is $\mathcal{O}(N_x^3)$ where $N_x$ denotes the number of collocation points in space. 
%The direct LU factorization used in the Newton step has $\mathcal{O}(N_x^3)$ computational complexity, but we found it quite robust. Also, the factorization costs are insignificant for the target problem sizes, $N_x=$ 200 to 400.
%The use of direct factorization is justified by the problem size (i.e., $N_x=200$ for medium resolution run, and $N_x=400$ for high resolution run).  

%Note that using Chebyshev collocation points naturally creates an adaptive grid in space where the higher resolution points are closer to the glow discharge walls. The above is useful to resolve the sheath region efficiently. 

\subsection{Hybrid fluid-BTE solver}
\label{subsec:pde_solver}
Here, we discuss the discretization of~\Cref{eq:glow_hybrid}. First, let us consider the velocity space discretization of \Cref{eq:hybrid_cnt_b}. The rate of change in the EDF due to the velocity space advection and collisions is given by \Cref{eq:bte-0d}. For the velocity space, we use spherical coordinates with a mixed Galerkin and collocation scheme for the angular direction and a Galerkin scheme for the radial coordinate discretization. We discretize $f(\vect{v}, t)$ as follows:
\begin{equation}
	f(\vect{v}, t) = \sum_{klm} f_{klm}\of{t} \Phi_{klm}\of{\vect{v}} \text{ where } \Phi_{klm}\of{\vect{v}}  = \underbrace{\phi_k\of{v}}_{\text{B-Spline basis}} \overbrace{Y_{lm}\of{v_\theta, v_\phi}}^{\tiny\text{spherical harmonics}}, \label{eq:f_expansion}
\end{equation} where $\biggl\{\phi_{k}\of{v}\biggr\}_{k=0}^{N_r-1}$ are cubic B-splines defined on a regular grid, and $Y_{lm}\of{\vtheta, \vphi}$ are defined as 
\begin{align}
	Y_{lm}\of{\vtheta,\vphi} &= U_{lm} P^{|m|}_l\of{\cos\of{\vtheta}} \alpha_m\of{\vphi} \text{ , } \nonumber \\
	U_{lm} = 
	\begin{cases}
		(-1)^m \sqrt{2} \sqrt{\frac{2l+1}{4\pi} \frac{(l-|m|)!}{(l+|m|)!}}, &m\neq 0, \\
		\sqrt{\frac{2l+1}{4\pi}}, &m = 0,
	\end{cases} \text{ , } &
	\alpha_m\of{\vphi}  =
	\begin{cases}
		\sin\of{|m|\phi}, &m < 0, \\
		1, &m = 0,\\
		\cos\of{m\phi}, &m > 0.
	\end{cases} \label{eq:sph_harmonics}
\end{align}
The $l$ index is also referred to as a polar mode, and the $m$ index as an azimuthal mode. Now, assume that $f_{klm}\of{t=0}=0,\  \forall m>0$. By aligning $\vect{E}$ to the velocity space z-axis, so that $\vect{E} = E \vect{\hat{e}_z} =  E \of{\cos(\vtheta) \vect{\hat{e}_r} - \sin(\vtheta)\vect{\hat{e}_\theta}}$ we ensure that the $\vect{E}$ acceleration excites only the polar modes~\Cref{eq:f_expansion}. This fact and the isotropic scattering assumption ensure that the BTE solutions preserve the azimuthal symmetry in the velocity space and thus $f_{klm}\of{t}$ remains zero $\forall t>0,\ m>0$. This essentially reduces the representation from 1D3V to 1D2V. Therefore, the EDF representation only requires $\{Y_{l0}\}_{l}$ modes. For this reason, we drop the azimuthal index $v_{\phi}$ in the remainder of the paper. Let $p$, $k$ denote the indices of the basis functions along the radial coordinate; $q$, $l$ denote the indices in the polar angle; and $\phi_p\of{v}$, $\phi_k\of{v}$ denote the test and trial B-splines in radial coordinate. Under these definitions, the discretized weak form of the collision operator is given by

\begin{multline}
    n_0 {[\vect C_{en}]}^{pq}_{kl} = n_0 \myint_{\reals^+} v^2 \sigma_T\of{v} \phi_{k}\of{v} \delta_{ql} \of{\phi_{p}\of{u} \delta_{q0}  -\phi_{p}\of{v}} \diff{v}, \\
    \text{ for } p,k \in \{0,\hdots, N_r-1\} \text{ and } q,l \in \{0, \hdots,  N_l-1\} \label{eq:c_en_mat_1d}.
\end{multline}
For non-zero heavy temperature, \Cref{eq:c_en_mat_1d} gets an additional correction term, given by
\begin{multline}
	n_0 T_0 {[\vect C_{T}]}^{pq}_{kl} = \frac{n_0 T_0 k_B}{m_0} \delta_{q0} \myint_{\reals^+} v^3 \sigma_T\of{v} \partial_v \phi_p\of{v} \partial_v \phi_k\of{v} \diff{v}, \\
	\text{ for } p,k \in \{0,\hdots, N_r-1\} \text{ and } q,l \in \{0, \hdots,  N_l-1\} \label{eq:c_T0_mat_1d}.
\end{multline} In \Cref{eq:c_en_mat_1d} and \Cref{eq:c_T0_mat_1d} $\sigma_T$ denotes the total collisional cross-section, $n_0$ and $T_0$ denotes the background neutral density and temperature. In our case, we have momentum transfer and ionization collisions. For multiple collisions, the effective collision operator is given by $\vect C_{en}^{\text{effec.}} = \sum_{i\in {\text{collisions}}}$ $n_i \vect{C_{en}} \of{\sigma_i} + n_0 T_0 \vect C_{T}\of{\sigma_0}$ where $0$ index denotes the ground state heavy species, and $\sigma_0$ denotes the total cross-section for the momentum transfer collisions. The discretized velocity space acceleration operator is given by
\begin{multline}
	\small
	[\vect A_v]^{pq}_{kl} = \int_{\reals^+}  
		\delta_{(q+1) l} v^2 \phi_p\of{v} \of{A_M\of{l} \partial_v\phi_k\of{v} + A_D\of{l}\frac{1}{v} \phi_k\of{v}} + \\
		\delta_{(q-1) l} v^2 \phi_p\of{v} \of{B_M\of{l} \partial_v\phi_k\of{v} + B_D\of{l}\frac{1}{v} \phi_k\of{v}} 
	 \diff{v}, \\
	 \text{ for } p,k \in \{0,\hdots, N_r-1\} \text{ and } q,l \in \{0, \hdots,  N_l-1\} \label{eq:adv_v_ws},
\end{multline} where $A_M$, $A_D$, $B_M$, and $B_D$ are coefficients defined by
\begin{equation}
A_M(l) = \frac{l}{\sqrt{4l^2-1}}\text{, } B_M(l) = \frac{l+1}{\sqrt{4l^2-1}}\text{, } A_D(l) = \frac{l^2}{\sqrt{4l^2-1}}\text{, } B_D(l) = \frac{l(l-1)}{\sqrt{4l^2-1}}. 	
\end{equation}
To summarize, the velocity space discretized BTE is given by
\begin{equation}
	\partial_t \of{\vect M_{v,\vtheta} \vect f} = \of{\vect{C}_{en} + E \vect A_v}\vect{f} \label{eq:discretized_bte_vspace},
\end{equation} where $\vect M_{v,\vtheta}$, $\vect C_{en}$, and $\vect A_v$ $\in \reals^{N_rN_l \times N_rN_l} $ denote the Galerkin mass matrix, electron-heavy collisions and the velocity space acceleration operators respectively. More details on the velocity space discretization can be found in \cite{fernando0DBTE}. 

Now we discuss the spatial discretization of \Cref{eq:hybrid_cnt_b}. We use a Chebyshev collocation scheme for the spatial discretization of~\Cref{eq:hybrid_cnt_b}. The spherical harmonic representation is inadequate to impose the $\vtheta$ discontinuous boundary conditions specified in~\Cref{eq:hybrid_cnt_b_bdy}. This is resolved by a mixed Galerkin and collocation representation in $\vtheta$. Let $\{x_i\}^{^{N_x}-1}_{i=0}$ and $\{\vtheta_a\}^{^{N_\vtheta}-1}_{a=0}$ be collocation points in $x$, and $\vtheta$ coordinates. With these collocation points, the EDF representation is given by
\begin{multline}
	f(x_i, v, \vtheta_a, t) = \sum_{k=0}^{N_r-1} f_{k}\of{x_i, \vtheta_a, t} \phi_{k}\of{v} = \sum_{k=0}^{N_r-1} f_{ika}\of{t} \phi_{k}\of{v}, \\ \text{ where } f_{ika}(t) \equiv f_k(x_i, \vtheta_a, t), 
\end{multline}
and the semi-discrete form of~\Cref{eq:hybrid_cnt_b} is given by 
\begin{equation}
\partial_t f(t, x, v, \vtheta_a) + v\cos\of{\vtheta_a} \partial_x f(t, x, v, \vtheta_a) = \of{\partial_t f}_{\text{$\vect{v}$-space}}\of{t, x, v, \vtheta_a} \label{eq:bte_1d_semi_discrete}.
\end{equation} Let $\vect{P}_{S}$ denotes the $\vtheta$ ordinates to spherical harmonics projection, and $\vect{P}_{O}$ denotes the spherical harmonics to $\vtheta$ ordinates projection operators. These operators are defined by
\begin{multline}
	\vect P_S = \vect I_v \otimes \vect T_S \text{ , } \vect P_O = \vect I_v \otimes \vect T_O \text{ where } 
	[\vect T_{S}]^{q}_{a} = Y_{q}\of{\vtheta_a} w_a \text{ and }\\
	[\vect T_{O}]^{a}_{q} = Y_{q}\of{\vtheta_a}, 
	\text{ for } q \in \{0, \hdots, N_l-1\}\text{, } a\in \{0, \hdots, N_\vtheta-1\}\label{eq:proj_ops} .
\end{multline} Here, $\otimes$ denotes the matrix Kronecker product, $\vect I_v\in \reals^{N_r\times N_r}$ denotes the identity matrix in $v$ coordinate, and $w_a$ denotes the quadrature weights for the spherical basis projection. We use a Chebyshev collocation method in space, a Galerkin discretization in the $v$ coordinate, and a mixed Galerkin and collocation scheme in the $\vtheta$ coordinate for the discretization of~\Cref{eq:bte_1d_semi_discrete}. Let $\vect{D}_x$ denotes the discrete Chebyshev-basis derivative operator and $\vect{A}_x$ denotes the $v$ coordinate Galerkin projection of the spatial advection term in~\Cref{eq:bte_1d_semi_discrete}. The $\vect{A}_x$ operator is given by
\begin{multline}
	\vect{A}_x = \vect{D}_{\vtheta} \otimes \vect{G}_{v} \text{ where,}\quad
	[\vect D_\vtheta]_{ab} = \delta_{ab} \cos\of{\vtheta_a} \text{ and } \\ [\vect{G}_{v}]_{pk} =\int_{\reals^+} v^3 \phi_{p}\of{v} \phi_{k}\of{v}\diff{v}, \\
	\text{ for } p,k \in \{0,\hdots, N_r-1\} \text{ and } a, b \in \{0, \hdots, N_{\vtheta}-1\}
	\label{eq:A_x}.
\end{multline} By putting everything together, with $x$, $v$, and $\vtheta$ discretized BTE is given by
\begin{multline}
   \partial_t \vect F + \vect A_x \vect F \vect D_x^T = \vect P_{O}\vect C_{en} \vect P_{S} \vect{F} + \vect P_O \vect A_v \vect P_{S} \of{\vect E \pdot \vect{F}} \text{ where } \\
   \vect F \in \reals^{N_r N_\vtheta \times N_x}\text{, }\vect{P}_s
   \in \reals^{N_rN_l \times N_r N_\vtheta}\text{, }\\
   \vect{C}_{en}\text{, }\vect{A}_v \in \reals^{N_r N_l \times N_r N_l}\text{, }
   \vect{P}_O \in \reals^{N_r N_\vtheta \times N_r N_l}\text{, }
   \vect{D}_x \in \reals^{N_x \times N_x}\text{, }\vect{A}_x\in \reals^{N_rN_\vtheta \times N_r N_\vtheta}, \\
   [\vect F]_{i,:} = f_{ika}\text{, } 
   \text{ for } i \in \{0, \hdots ,N_x-1\} \text{, } k \in \{0, \hdots N_r-1\} \text{, and } a \in \{0, \hdots N_\vtheta-1\}    \label{eq:bte_1d_discrete}.
\end{multline} Here, $\vect{E} \in \reals^{N_x}$ and $\pdot$ denotes the column wise element product. Also, it is important to note that the operators in \Cref{eq:bte_1d_discrete} is properly scaled by Galerkin mass matrices, i.e., from here on, $\vect{G}_v \equiv \vect{M}^{-1}_v \vect{G}_v$, $\vect{A}_x \equiv \vect{D}_\vtheta \otimes \vect G_v$, $\vect{C}_{en} \equiv \vect M^{-1}_{v,\vtheta} \vect C_{en}$, and $\vect{A}_v \equiv M^{-1}_{v,\vtheta} \vect A_v$, where $\vect{M}_v$ and $\vect{M}_{v,\vtheta}$ denote the standard Galerkin mass matrices~\cite{ciarlet2002finite} in $v$ and $v, \vtheta $ coordinates respectively.  \Cref{eq:hybrid_cnt_a,eq:hybrid_cnt_c} are discretized with the same Chebyshev collocation scheme we used for~\Cref{eq:hybrid_cnt_b}. Let $\vect{k}_i\in \reals^{N_r N_\vtheta}$ denotes the discretized ionization rate coefficient operator, $\vect{u}\in\reals^{N_r N_\vtheta}$ denotes the zeroth-order moment operator where $\vect{n}_e = \vect{u}^T \vect{F}$, and $\vect{L}\in\reals^{N_x \times N_x}$ denotes the discretized electrostatics operator where $\vect{E}=\vect{L}\of{\vect{n}_i-\vect{n}_e}$. Then, the discretized hybrid model is given by
\begin{subequations}
	\begin{align}
		&\partial_t \vect n_i  + \vect D_x \vect J_i \of{\vect E, \vect n_i} = \of{\vect{k}^T_i \vect F} n_0 \vect n_e \label{eq:hybrid_discrete_a},\\
		&\partial_t \vect F + \vect A_x \vect F \vect D_x^T = \vect P_{O}\vect C_{en} \vect P_{S} \vect{F}  +  \vect P_O \vect A_v \vect P_{S} \of{\vect E \pdot \vect{F}}   \label{eq:hybrid_discrete_b},\\
		&\vect{E} = \vect{L}\of{\vect n_i - \vect n_e}, \text{ where } \vect n_e = \vect u^T \vect F \label{eq:hybrid_discrete_c}.
	\end{align}\label{eq:hybrid_discrete}
\end{subequations}The total number of unknowns for the discretized system is $N_x(1 + N_{r} N_{\vtheta})$ where $N_x$, $N_r$, and $N_{\vtheta}$ denote the number of Chebyshev collocation points in $x$, the number of B-spline basis used in $v$, and the number of collocation points in $\vtheta$.

\subsubsection{Timescales analysis}
\label{subsubsec:ts_analysis}
Here, we present timescales analysis for \Cref{eq:hybrid_discrete}. \Cref{tab:ts_analysis}~summarizes the main ion and electron timescales for a RF-GDP with $p_0=1$ Torr. The input voltage oscillation period is set to $7.374 \times 10^{-8}$ s for all RF-GDP cases considered in this paper. The timescale for the collision operator is given by
\begin{multline}
	\Delta t \frac{\norm{\vect{C}_{en} \vect{F}}}{\norm{\vect{F}}} \leq \Delta t \norm{\vect{C}_{en}} \leq \Delta t n_0\max\of{\norm{\vect{v}}\sigma_{0}, \norm{\vect{v}}\sigma_{i}}\leq \frac{\norm{\Delta \vect{F}}}{\norm{\vect{F}}}\\ \implies
	\Delta t \leq \frac{\norm{\Delta \vect{F}}}{\norm{\vect{F}}} \frac{1}{n_0\max\of{\norm{\vect{v}}\sigma_{0}, \norm{\vect{v}}\sigma_{i}}}= \frac{\norm{\Delta \vect{F}}}{\norm{\vect{F}}} \frac{1}{n_0\max\of{\norm{\vect{v}}\sigma_{0}}} \label{eq:coll_timescale}.
\end{multline} The reaction and collision terms determine the fastest timescale for ions and electrons. However, as shown in \Cref{tab:ts_analysis}, ions have much slower timescales than electrons. Therefore, the electron timescale determines the electrostatic timescale. A change in the electron number density $n_e(\vect{x}, t)$ is caused by the BTE spatial advection and reaction terms. While the velocity space advection is not directly coupled to $n_e(\vect{x}, t)$, it is indirectly coupled through the collision term. This is due to the EDF tails being populated due to the velocity space advection, and high-energy electrons are likely to trigger ionization collisions. Therefore, the electrostatic and the BTE coupling timescale is governed by the position-velocity space advection and the reaction term timescales. The collisional timescale gives the relative change in the EDF due to collisions. Typically, Since $\sigma_0 \geq \sigma_{\text{i}}$, the collisional timescale is determined by the momentum transfer cross-section. 
%Operator split schemes that decouple position space and velocity space operators

\begin{table}[tbhp]
	\centering
	\resizebox{\textwidth}{!}{
		\begin{tabular}{||c|c|c||}
			\hline 
			Species  & Term & $\Delta t$ [s]\\
			\hline
			\multirow{3}{*}{Ions} 	& $\nabla_{\vect{x}} \cdot \of{\mu_i \vect{E} n_i}$ -- advection 		 & $\frac{\Delta x}{\mu_i \norm{\vect{E}_{\text{max}}}} \approx \mathcal{O}(10^{-9})$  \\		[0.1cm]
			& $\nabla_{\vect{x}} \cdot \of{D_i \nabla_{\vect{x}} n_i}$ -- diffusion  & $\frac{\Delta x^2}{ 2 D_i } \approx \mathcal{O}(10^{-7}) $  \\ 	[0.1cm]
			& $k_i n_0 n_e$ -- reaction   											 & $\frac{\Delta n_i}{n_i}\frac{n_i}{k_{i,\text{max}} n_0 n_e} \approx \mathcal{O}(10^{-10})$   \\ [0.1cm]
			\cline{2-3}
			\multirow{3}{*}{Electron BTE}  & $\vect{v} \cdot \nabla_{\vect{x}} f$ -- $\vect{x}$-advection 		 				 & $\frac{\Delta x}{\norm{\vect{v}}} \approx \mathcal{O}(10^{-12})$\\ [0.1cm]
			& $\frac{q_e \vect{E}}{m_e} \cdot \nabla_{\vect{v}} f$ -- $\vect{v}$-advection 		 & $\frac{\norm{\Delta \vect v} m_e}{q_e\norm{\vect{E}_{\text{max}}}} \approx \mathcal{O}(10^{-12})$\\ 	[0.1cm]
			&  $k_i n_0 n_e = n_0 \int_{\vect{v}} \norm{\vect v}^3 \sigma_{i}\of{\norm{\vect v}} f\of{\vect{v}} \diff{\vect{v}}$ -- reaction & $\frac{\Delta n_e}{n_e}\frac{1}{k_{i,\text{max}} n_0 } \approx \mathcal{O}(10^{-12})$ \\ [0.1cm]
			&$ \vect{C}_{en}\vect{F}$ -- collisions  & $\frac{\norm{\Delta \vect{F}}}{\norm{\vect{F}}} \frac{1}{n_0 \max\of{ \norm{\vect{v}} \sigma_0}} \approx \mathcal{O}(10^{-13})$\\ [0.2cm]
			\cline{2-3}
			\multirow{2}{*}{Driving voltage} & &\\[0.01cm]
											 & $V(t) = V_0 \sin\of{2\pi \zeta t}$ & $1/\zeta=7.374 \times 10^{-8}$ \\ [0.1cm]
			\hline
	\end{tabular}}
	\caption{A summary of the timescales for advection, diffusion, and reaction terms for the hybrid formulation, assuming an explicit time integration scheme.  Here, we assume $\frac{\Delta x}{L}\approx \mathcal{O}(10^{-3})$, $\frac{\norm{\Delta \vect{v}}}{\norm{\vect v}} \approx \mathcal{O}(10^{-3})$, $\frac{\norm{\Delta\vect{F}}}{\norm{\vect{F}}} \approx \mathcal{O}(10^{-3})$, and $\frac{\Delta n_e}{n_e} \approx \mathcal{O}(10^{-3})$. The maximum absolute quantities occurs closer to the electrodes with $\norm{\vect E_{\text{max}}}=8\times10^{4}$ [V/m], $k_{i,\text{max}}=10^{-14}$ [$m^3/s$], and the corresponding ionization degree $\frac{n_i}{n_e}\approx\mathcal{O}(10^2)$. For the other transport coefficients, we use the values summarized in \Cref{t:model_parameters} with the $p_0=$ 1 Torr case. \label{tab:ts_analysis}}
\end{table}
%Considering the plasma oscillation frequencies for electrons $\omega_{e} = \sqrt{q_e^2 n_e /m_e/\epsilon_0}$ and ions $\omega_{i} = \sqrt{q_e^2 n_i /m_i/\epsilon_0}$ we can write $\omega_{e} / \omega_{i} = \of{m_i / m_e}^{1/2} \of{n_e / n_i}^{1/2}$. For RF-GDPs, typically $n_e \sim n_i \sim 10^{17}$ $m^{-3}$, $f_{e} \approx 8.9 \times 10^{9}$ $s^{-1}$ and $f_{i} \approx 3.3 \times 10^{7}$ $s^{-1}$. Therefore, electrons have a higher plasma frequency than ions. This is primarily due to the small mass ratio between electrons and ions, $m_e/m_i = 1.36\times 10^{-5}$. 

\subsubsection{Hybrid solver time integration}
Typically, we want to evolve the system to the order of thousand cycles or $10^{-5}$ seconds, and require ten million timesteps. An implicit scheme with an iterative solve is possible as the matrix-vector products can be also done in an optimal way, but designing a good preconditioner is not trivial. Instead we opt for an operator split scheme.
As we can see in~\Cref{tab:ts_analysis}, the ions have slower timescales compared to electrons. Hence, we use an operator split scheme to decouple ion and electron transport equations. Since electrons determine the electrostatics coupling timescale, \Cref{eq:hybrid_discrete_a} is solved with decoupled electrostatics.
%The ion and electron plasma frequencies determine the $\vect{E}$ field coupling time scales for \Cref{eq:hybrid_discrete_a,eq:hybrid_discrete_b}. This is mainly due to the electron-to-ion mass ratio. Therefore, ions respond slowly compared to the electrons for a given perturbation in the electric field. Hence, \Cref{eq:hybrid_discrete_a} is solved with frozen $\vect{E}$ and $\vect{F}$ values. 

\par \textbf{Heavies evolution}: For a given state $(\vect n_i^n, \vect F^n)$ at time $t=t_n$, the ion transport equation is solved implicitly with lagged $\vect F^n$ and $\vect E^n = \vect L (\vect n^n_i - \vect u^T \vect F^n)$. This coupling results in a linear system to be solved for the ions state at time $t_{n+1}=t_n +  \Delta t$, where $\Delta t$ denotes the time step size. So, the ions update in our operator-split scheme is given by
\begin{equation}
	\frac{\vect n^{n+1}_i -\vect n^n_i}{\Delta t} + \vect{D}_x \vect J_i \of{\vect E^n, \vect n^{n+1}_i}= \of{\vect k_i^T \vect F^n} n_0 \of{\vect u^T \vect F^n} \label{eq:fluid_step}.
\end{equation}

%The electrons are strongly coupled to the electric field through~\Cref{eq:hybrid_discrete_c}. Therefore, \Cref{eq:hybrid_discrete_b,eq:hybrid_discrete_c} are solved together.
%to account for the $\vect{E}$ field variation due to electron transport.

\par \textbf{Electron evolution}: We consider the evolution of \Cref{eq:hybrid_discrete_b,eq:hybrid_discrete_c} assuming a fixed $\vect{n}_i$. We consider two alternative schemes for the electrons: a semi-implicit and a fully-implicit one. The semi-implicit scheme compute the EDF at time $t + \Delta t$ is given by
\begin{subequations}
	\begin{empheq}[left =\text{\small semi-implicit} \empheqlbrace]{align}
        &\frac{\vect{F}^{n+1/2} - \vect{F}^n}{\Delta t/2} =  -\vect A_x \vect F^{n+1/2} \vect D_x^T, t \in \of{t_n , t_n + \frac{\Delta t}{2}} \label{eq:bte_semi_implicit_xspace0}, \\
        &\vect{E}^{1/2}=\vect L \of{\vect n_i - \vect u^T \vect F^{n + 1/2}} \label{eq:bte_semi_implicit_efield},\\
        & \frac{\vect{F}^{*} - \vect{F}^{n+1/2}}{\Delta t} = \vect P_{O}\vect C_{en}\vect P_{S} \vect{F}^*  + \nonumber \\
        &\qquad \qquad \vect P_O \vect A_v \vect P_{S} \of{\vect{E}^{1/2} \pdot \vect{F}^*}, t \in \of{t_n, t_n + \Delta t}, \label{eq:bte_semi_implicit_vspace}\\
        &\frac{\vect{F}^{n+1} - \vect{F}^*}{\Delta t / 2} =  -\vect A_x \vect F^{n+1} \vect D_x^T, t \in \of{t_{n+1/2} , t_{n+1/2} + \frac{\Delta t}{2}} \label{eq:bte_semi_implicit_xspace1}.
    \end{empheq} \label{eq:bte_semi_implicit}
\end{subequations}
The overview of the semi-implicit scheme is summarized below. 
\begin{itemize}
\item \textbf{Spatial advection}: \Cref{eq:bte_semi_implicit_xspace0,eq:bte_semi_implicit_xspace1} corresponds to the spatial advection of electrons. We use the the eigen decomposition of $\vect{G}_v=\vect{U}\vect{\Lambda}U^{-1}$ to diagonalize \Cref{eq:bte_semi_implicit_xspace0,eq:bte_semi_implicit_xspace1}. Since $\vect{G}_v$ is a product of two positive definite matrices, it has real positive eigenvalues. Upon diagonalization, \Cref{eq:bte_semi_implicit_xspace0,eq:bte_semi_implicit_xspace1} become series of decoupled $N_rN_\vtheta$ spatial advection equations where the advection velocity is given by $\vect{a} = \Lambda_i \cos(\vtheta_j) \vect{\hat{e}}_x$ for $i=\{0,\hdots N_r-1\}$ and $j=\{0,\hdots N_\vtheta-1\}$ where $N_r$ and $N_\vtheta$ denotes the number of B-splines and $\vtheta$ ordinates used. For each $i$ and $j$, we precompute and store the inverted $\left(\vect{I}_x + \frac{1}{2} \Delta t \Lambda_i \cos(\vtheta_j) \vect D_x \right)^{-1} = \vect Q_{ij}$ operators and use them for spatial advection. Here, $\vect{I}_x\in \reals^{N_x\times N_x}$ denotes the identity matrix in the $x$ dimension. This direct inversion is performed only once and reused throughout the time evolution. The storage cost for the inverted matrices is given by $\mathcal{O}(N_r N_\vtheta N_x^2)$. \Cref{alg:bte_spatial_adv} presents the overview of the BTE spatial advection step. In the implementation, the application of $\vect Q_{ij}$ is performed as a batched general matrix-matrix multiplication (GEMM).
\item \textbf{Electrostatics}: \Cref{eq:bte_semi_implicit_efield} computes the updated electric field $\vect{E}^{1/2}$ following the spatial advection. Here, $\vect{L}\in \reals^{N_x \times N_x}$ denotes the discretized $\nabla_{\vect{x}} \Delta^{-1}_{\vect x}$ operator which is computed once and reused throughout the time evolution. 
\item \textbf{Velocity space update}: We use the generalized minimal residual method (GMRES) to solve~\Cref{eq:bte_semi_implicit_vspace}. In~\Cref{eq:bte_semi_implicit_vspace}, the discretized EDF is spatially decoupled, and the velocity space left-hand side operator varies spatially and is given by
\begin{multline}
	\biggl(\vect{I}_{v,\vtheta} -\Delta t \vect P_{O}\of{\vect C_{en}  + \bigl[\vect{E}^{1/2}\bigr]_{i} \vect A_v  }  \vect P_{S}\biggr) \vect F^* = \vect F^{n + 1/2}  \\ \text{ where } \vect E^{1/2} \in \reals^{N_x}, 
	\text{ for } i \in \{0,\hdots, N_x - 1\} \label{eq:vspace_batched_system}.
\end{multline} Here, $\vect{I}_{v,\vtheta}$ denotes the identity operator in the velocity space, and $[\vect{E}^{1/2}]_i$ denotes the electric field evaluated at the $i$-th spatial collocation point. Therefore, the solve of~\Cref{eq:bte_semi_implicit_vspace} reduces to a $N_x$ decoupled linear systems. For the GMRES solve, these linear systems are solved as a batched system. The batched right-hand side evaluation is summarized in \Cref{alg:vspace_action}. We use a domain decomposition preconditioner for the GMRES solve. Let $E_0$ denotes an upper bound on the electric field such that $\norm{\vect{E}\of{\vect x, t}} \leq E_0 \forall t$ and $-E_0 = S_0 < S_1 < \hdots < S_{N_c}=E_0$ be a time independent non-overlapping partition of $(-E_0, E_0)$. We precompute a series of operators $\{H_p\}_{p=1}^{N_c}$ where each $\vect H_p\in \reals^{N_r N_l}$ is given by
\begin{equation}
	\vect H_p = \of{\vect I -\Delta t \of {\vect C_{en} + \frac{1}{2}\of{S_{p-1} + S_{p}} \vect A_v}}^{-1} 
	\text{ for } p\in\{1,\hdots N_c\}.
\end{equation} Here, $\vect I\in \reals^{N_r N_l}$ denotes the identity matrix. For a specified $\vect{E} \in \reals^{N_x}$, we define $N_c$ partitions where partition $w_k$ is given by $w_k=\{i\quad | \  S_{k-1} \leq [\vect{E}]_i < S_{k}\}$ for $k\in \{1,\hdots,N_c\}$. We refer to this as the preconditioner setup step, which is performed once per each timestep. The precondition operator evaluation is summarized in \Cref{alg:vspace_action_precon}.
%$\{\vect H_p\}_{p=1}^{N_c}$
\end{itemize}

\begin{algorithm}[!tbhp]
	\begin{algorithmic}[1]
		\Require $\vect{F}^n\in \reals^{N_rN_\vtheta \times N_x}$ -- EDF at time $t_n$, \nonumber\\
		$\biggl\{\vect{Q}_{ij} = \left(\vect{I}_x + \frac{1}{2} \Delta t \Lambda_i \cos(\vtheta_j) \vect D_x \right)^{-1}\biggr\}_{(i,j)\in \{0,..,N_r-1\}\times\{0,..,N_\vtheta-1\}}$ -- inverted spatial advection operators, and $\vect{G}_v = \vect{U} \vect \Lambda \vect U^{-1}$
		\Ensure Solution at time $t_n + \Delta t/2$, $\vect{F}^{1/2} \in \reals^{N_r N_\vtheta \times N_x}$ 
		\State $\vect Y \leftarrow \text{unfold}(\vect F, (N_r, N_\vtheta, N_x))$ \Comment{$\vect Y \in \reals^{N_r \times N_\vtheta\times N_x}$}
		\State $[\vect Y]_{ijk} \leftarrow [\vect U^{-1}]_{im} \times_{m} [\vect F^n]_{mjk}$ \Comment{contraction on $m$ index}
		\For{$i \in \{0,\hdots, N_r-1\}$}
		\For{$j \in \{0,\hdots, N_\vtheta-1\}$}
		\rlap{\smash{$\left.\begin{array}{@{}c@{}}\\{}\\{}\end{array}\color{black}\right\}%
				\color{black}\begin{tabular}{l} Batched GEMM kernels \end{tabular}$}}
		\State $\vect Y_{ijk} \leftarrow \vect Q_{ijkl} \times_l \vect Y_{ijl}$ \Comment{$\vect Q_{ij}\in \reals^{N_x \times N_x}$ and contraction on $l$ index}
		\EndFor
		\EndFor
		\State $[\vect Y]_{ijk} \leftarrow [\vect U]_{im} \times_{m} [\vect F^n]_{mjk}$ \Comment{contraction on $m$ index}
		\State $\vect F^{1/2} \leftarrow \text{fold}(\vect Y , (N_rN_\vtheta \times N_x))$ \Comment{$\vect F^{1/2} \in \reals^{N_rN_\vtheta \times N_x}$}
		\State \Return $\vect{F}^{1/2}$
	\end{algorithmic}
	\caption{BTE spatial advection. \label{alg:bte_spatial_adv}}
\end{algorithm}
\begin{algorithm}
	\begin{algorithmic}[1]
		\Require $\vect F \in \reals^{N_r N_\vtheta \times N_x}$, $\vect E \in \reals^N_x$ -- electric field, $\vect C_en$, $\vect A_v$, $\vect P_O$, $\vect P_S$, and $\Delta t$
		\Ensure $\biggl(\vect{I}_{v,\vtheta} -\Delta t \vect P_{O}\of{\vect C_{en}  + \bigl[\vect{E}\bigr]_{i} \vect A_v  }  \vect P_{S}\biggr) \vect F$ -- operator action on $\vect F$
		\State $\vect F_S \leftarrow \vect P_S \vect F$
		\State $\vect F_C \leftarrow \vect C_{en} \vect F_s$ \quad and \quad $\vect F_A \leftarrow \vect A_v \vect F_s$
		\State $\vect G  \leftarrow  \vect F - \Delta t \vect P_O \of{\vect F_C + \vect E \pdot \vect F_A}$
		\State \Return $\vect G$
	\end{algorithmic}
	\caption{BTE velocity space operator action. \label{alg:vspace_action}}
\end{algorithm}
\begin{algorithm}
	\begin{algorithmic}[1]
		\Require $\vect F $, $\{\vect H_p\}_{p=1}^{N_c}$ -- precondition operators, $\{w_p\}_{p=1}^{N_c}$ -- domain decomposition, $\vect P_S $, $\vect P_O$
		\Ensure Preconditioner action $: \reals^{N_r N_\vtheta \times N_x} \rightarrow \reals^{N_r N_\vtheta \times N_x}$
		\State $\vect G \leftarrow \vect 0 $ \Comment{$\vect G \in \reals^{NrN_l \times N_x}$}
		\State $\vect F_S \leftarrow \vect P_S \vect F$ \Comment{$\vect F_s \in \reals^{NrN_l \times N_x}$}
		\For {$p \in \{1,\hdots, N_c\}$ }
			\State $\vect G[:, w_p] \leftarrow \vect H_p \vect F_S[:, w_p]$
		\EndFor
		\State $\vect G \leftarrow \vect P_O \vect G$
		\State \Return $\vect G$ \Comment{$\vect G \in \reals^{N_r N_\vtheta \times N_x}$}
	\end{algorithmic}
	\caption{BTE velocity space GMRES preconditioner. \label{alg:vspace_action_precon}}
\end{algorithm}

The semi-implicit approach has a timestep size restriction that arises from the decoupled electrostatics, spatial, and velocity space solves. This is determined by the BTE spatial advection and reaction timescales (see \Cref{subsubsec:ts_analysis}). To address this timestep size restriction, we propose a fully-implicit scheme for the BTE, which is given by 
\begin{subequations}
	\begin{empheq}[left =\text{\small fully-implicit} \empheqlbrace]{align}
		\frac{\vect F^{n+1} - \vect F^n}{\Delta t}  &= -\vect A_x \vect F^{n+1} \vect D_x^T + \vect P_{O}\vect C_{en} \vect P_{S} \vect{F}^{n+1} + \nonumber \\
		& \qquad\qquad \vect P_O \vect A_v \vect P_S \of {\vect E(\vect F^{n+1}) \pdot \vect F^{n+1}}     \label{eq:bte_fully_implicit_a}, \\
		\vect{E}\of{\vect{F}} &= \vect L \of{\vect n_i - \vect u^T \vect F} \label{eq:bte_fully_implicit_b}.
	\end{empheq} \label{eq:bte_fully_implicit}
\end{subequations} 
We use Newton's method  with line search to solve the nonlinear system by~\Cref{eq:bte_fully_implicit}. The application of the Jacobian, evaluated at $\vect F$, on a given input $\vect X$ is given by
\begin{multline}
	%\vect{J_F} \vect X = -\vect A_x \vect X \vect D_x^T + \vect P_O \of{ \vect C_{en} + \vect E\of{\vect F} \pdot \vect A_v} \vect P_S \vect X  +  \vect P_O \of{\vect E \of{\vect X} \pdot \vect A_v }\vect P_S \vect F \label{eq:1dbte_jac_action}.
	\vect{J_F} \vect X = -\vect A_x \vect X \vect D_x^T + \vect P_O \vect C_{en} \vect P_S \vect X  + \\ \vect P_O \vect A_v \vect P_S \of{E\of{\vect F} \pdot \vect X}  +  \vect P_O \vect A_v \vect P_S \of{\vect E \of{\vect X} \pdot \vect F} \label{eq:1dbte_jac_action}.
\end{multline}
Computing the linear step is, however, a challenge. If we try to assemble the Jacobian, we end up with a dense phase space operator with a prohibitive storage cost. Instead, we use a matrix-free GMRES solver, but it turns out the Jacobian is highly ill-conditioned. The solution is to use preconditioning. For this, we propose to use the linear operator corresponding to the update of the semi-implicit scheme defined in~\Cref{eq:bte_semi_implicit} as a preconditioner. We discuss the efficiency of the solver in the results section.

%This is a challenging system to solve. The key challenges include fully coupled phase space solve, operator assembly and storage cost is expensive, and the BTE equation becomes non-linear due to the $\vect{E}$ field coupling. Also, if we were to use an LU factorization for the Newton step, the cost would be $\mathcal{O}\of{N_r^3 N_x^3 N_{\vtheta}^3}$ which is infeasible for the problem sizes we consider. To mitigate these challenges, we use a matrix-free Newton's method to evolve~\Cref{eq:bte_fully_implicit} where the Jacobian action is given by
%We use GMRES to perform the Jacobian solve for the Newton iteration. We use~\Cref{eq:bte_semi_implicit} as a preconditioner for the GMRES solve of~\Cref{eq:bte_fully_implicit}. 
%We use an operator split similar to semi-implicit scheme as a preconditioner for the fully-implicit solve. 

\subsubsection{Hybrid solver complexity}
Recall that $N_x$, $N_r$, and $N_{\vtheta}$ denote the number of Chebyshev collocation points in position space, B-spline basis functions in $v$, and discrete ordinate points in $\vtheta$. This will result in $N_x(1 + N_r N_{\vtheta})$ degrees of freedom for the hybrid solver. Recall, $N_l$ denotes the number of spherical harmonics used for the mixed Galerkin and collocation discretization scheme used in $\vtheta$ coordinate.The velocity space operators $\vect{C}_{en}$, $\vect{A}_v$ are stored as dense matrices. Operators $\vect{A}_x$, $\vect P_S$, $\vect P_O$ are not assembled, and their actions are performed with tensor contractions with the stored $\vect D_\vtheta$, $\vect{G}_v$, $\vect T_S$, and $\vect T_O$ operators. These operators are stored as dense matrices. \Cref{tab:hybrid_comp_complexity}~summarizes all discretized BTE operators and their dimensions. It also provides the storage complexities for these operators and the computational complexity for their actions.
%As mentioned, the $\vect{v}$-space advection-collision given by~\Cref{eq:bte_vspace} is discretized using $N_l$ spherical harmonics in $\vtheta$. Therefore, this appears in the $\vect{A}_v$ and $\vect C_{en}$ costs. \Cref{tab:hybrid_comp_complexity}~summarizes complexity for storage and action of key BTE operators. 
\begin{table}[!tbhp]
	\centering
	\resizebox{\textwidth}{!}{
	\begin{tabular}{||c|c|c|c|c||}
		\hline
		Operator & Dimensions &  Storage & Action on $\small \vect{X} \in \reals^{N_r\times N_\vtheta \times N_x}$, $\vect X_S \in \reals^{N_r\times N_l \times N_x}$ & Complexity\\
		\hline
		%$\vect{F}$ -- degrees of freedom & $\mathcal{O} (N_x(1+N_rN_{\vtheta}))$ & --\\
		$\vect{D}_x$ & $N_x \times N_x $  & $\mathcal{O} (N_x^2)$ & $[\vect Y]_{ijk} = [\vect D_x]_{im} \times_m [\vect X]_{jkm}$  & $\mathcal{O}\of{N_x^2 N_r N_{\vtheta}}$ \\ [0.05cm]
		$\vect{A}_x=\vect D_\vtheta \otimes \vect G_v$ & $N_rN_\vtheta \times N_r N_\vtheta $  & $\mathcal{O} (N_r^2 + N_{\vtheta})$ & $[\vect Y]_{ijk} = \vect [D_\vtheta]_{jj} \pdot_j [\vect G_v]_{im} \times_m [\vect X]_{mjk}$ & $\mathcal{O} \of{N_x N_{\vtheta} N_r^2}$ \\ [0.05cm]
		$\vect P_S=\vect I_v \otimes \vect T_S$ & $N_r N_l \times N_v N_\vtheta$ & $\mathcal{O}(N_{\vtheta} N_l)$ & $[\vect Y]_{ijk} = [\vect T_S]_{jm} \times_m [\vect X]_{imk}$ & $\mathcal{O}\of{N_x N_r N_l N_\vtheta}$ \\ [0.05cm]
		$\vect P_O=\vect I_v \otimes \vect T_O$ & $N_r N_\vtheta \times N_r N_l$ & $\mathcal{O}(N_{\vtheta} N_l)$ & $[\vect Y]_{ijk} = [\vect T_O]_{jm} \times_m [\vect X_S]_{imk}$ & $\mathcal{O}\of{N_x N_r N_l N_\vtheta}$ \\ [0.05cm]
		$\vect{C}_{en}$ & $N_r N_l \times N_r N_l$ & $\mathcal{O}(N_r^2 N_l^2)$ & $[\vect Y]_{ijk} = [\vect C_{en}]_{ijlm} \times_{lm} [\vect X_S]_{lmk}$ & $\mathcal{O}\of{N_x N_r^2 N_l^2}$ \\ [0.05cm]
		$\vect{A}_{v}$ & $N_r N_l \times N_r N_l$ & $\mathcal{O} (N_r^2 N_l^2)$ & $[\vect Y]_{ijk} = [\vect A_{v}]_{ijlm} \times_{lm} [\vect X_S]_{lmk}$ & $\mathcal{O}\of{N_x N_r^2 N_l^2}$ \\ [0.05cm]
		\hline
	\end{tabular}}
\caption{Summary of storage and computational complexity for the discretized BTE operators. Here $\times_k$ denotes the contraction along index $k$, and $\pdot_k$ denotes element wise product along index $k$. \label{tab:hybrid_comp_complexity}}
\end{table} 

The computational complexities for $\vect A_x \vect F \vect D_x^T$ evaluation is given by $T_{\vect{x}\text{-rhs}} = \mathcal{O}\big(N_r N_{\vtheta} N_x \of{N_r + N_x}\big)$,  and the spatial advection solve in \Cref{alg:bte_spatial_adv} is $T_{\vect{x}\text{-solve}} = \mathcal{O}\of{N_r N_{\vtheta} N_x \of{N_r + N_x}}$. The computation cost for $P_O\of{\vect C_{en} + \vect E\pdot \vect A_v} P_S$ is $T_{\vect{v}\text{-rhs}} = \mathcal{O} \of{\of{2 N_\vtheta + N_r N_l} N_r N_l N_x}$. Application of the preconditioner has a similar costs so that $T_{\vect{v}\text{-precond}} = T_{\vect{v}\text{-rhs}}$. Hence, the overall computational cost for a single BTE time-step for the semi-implicit scheme is given by 
\begin{equation}
	T_{\text{semi-implicit}} =  T_{\vect{x}\text{-solve}} + k(T_{\vect{v}\text{-precond}} + T_{\vect{v}\text{-rhs}}) = T_{\vect{x}\text{-solve}} + 2kT_{\vect{v}\text{-rhs}} \label{eq:semi_imp_ts_cost}.
\end{equation} 
Here, $2kT_{\vect{v}\text{-rhs}}$ denotes the velocity space GMRES solver cost, and $k$ denotes the GMRES iterations for convergence. 
For the fully-implicit scheme, the right-hand side evaluation cost of the Jacobian action given in~\Cref{eq:1dbte_jac_action} is $T_{\vect{vx}\text{-jac}} = T_{\vect{x}\text{-rhs}} + T_{\vect{v}\text{-rhs}} + T_{\vect{E}\text{-solve}}$. Here, $\vect E$ solve cost is $T_{\vect{E}\text{-solve}} = \mathcal{O}(N_r N_\vtheta N_x + N_x^2)$, and since $(T_{\vect{x}\text{-rhs}} + T_{\vect{v}\text{-rhs}}) >> T_{\vect E\text{-solve}}$ it is omitted in the analysis. The operator split preconditioning cost for the fully implicit scheme is $T_{\vect{vx}\text{-precond}} = T_{\vect{x}\text{-rhs}} + T_{\vect{v}\text{-rhs}}$. Taken together, the computational complexity for a single timestep solve of the fully-implicit scheme is given by
\begin{equation}
T_{\text{fully-implicit}} = k(T_{\vect{vx}\text{-precond}} + T_{\vect{vx}\text{-jac}}) = 2k(T_{\vect{x}\text{-rhs}} + T_{\vect{v}\text{-rhs}}) \label{eq:bte_imp_complexity}.
\end{equation} Here, $k$ denotes the total Newton-GMRES iterations for convergence.
Let $(\Delta t_F,k_F)$ and $(\Delta t_S,k_S)$ tuples denote the timestep size and the number of iterations for the described fully-implicit and semi-implicit schemes respectively. The cost ratio $S$ of the semi-implicit to fully-implicit scheme is given by
\begin{equation}
	\frac{\Delta t_F \of{T_{\vect{x}\text{-rhs}} + 2k_S T_{\vect{v}\text{-rhs}}}}{\Delta t_S 2k_F \of{T_{\vect{x}\text{-rhs}} + T_{\vect{v}\text{-rhs}}}} \label{eq:ts_efficiency}.
\end{equation} 
Due to the nonlinearity and ill-conditioning of~\Cref{eq:bte_fully_implicit} it is hard to determine the effectiveness of the fully-implicit scheme. It depends on the preconditioned linear solves and the underlying prefactors in the complexity estimates, some of which are related to memory accesses and hardware performance. Next, we describe a series of numerical experiments to compare the two schemes. %Typically, we need $\Delta t_S \sim \frac{1}{\omega_{e}} < \Delta t_F$.% and the fully-implicit scheme is more efficient than the semi-implicit scheme, when $S > 1$.

\subsection{Implementation}
\label{subsec:implement_details}
We have implemented all the presented algorithms in Python and we release them in a library we call~\bte. We use \texttt{NumPy} and \texttt{CuPy} libraries for all the linear algebra operations. The above libraries provide an interface to basic linear algebra subprograms (BLAS) for CPU and GPU architectures. Both hybrid and the fluid model time integration supports GPU acceleration with \texttt{CuPy}. We use \texttt{CuPy} GMRES solver with \texttt{LinearOperator} class to specify operator and preconditioner actions. The batched GEMM operations are performed using the \texttt{einsum} function. \bte~is available at \url{https://github.com/ut-padas/boltzmann.git}.
\section{Results} \label{sec:results}
% \begin{itemize}
%     \item self-convergence tests (1D bte with specified E field)
%     \item cross-verification results with PIC-dsmc
%     \item glow discharge results
%     \begin{itemize}
%     	\item parametric study fluid and hybrid models under different pressure values. 
%		%\item can we describe the differences with drift diffusion approx. differences
%     \end{itemize}
% \end{itemize}
We organize the results as follows. \Cref{subsec:results_specified_E} presents a self-convergence study for the developed Eulerian BTE solver and cross-verification comparison with an in-house PIC-DSMC code. Performance evaluation of the developed algorithms is presented in~\Cref{subsec:peformance_glowd}. \Cref{subsec:results_glowd} presents a detailed comparison study between the hybrid and the fluid approximations of RF-GDPs for different pressure regimes. 

\subsection{Boltzmann transport with specified electric field}
\label{subsec:results_specified_E}

\begin{figure}[!tbhp]
	\begin{center}
		\subfloat[\hspace{-0.7in} \label{fig:case1_1_1}]{
			\begin{tikzpicture}
				\begin{semilogyaxis}[xlabel={$\hat{x}$}, ylabel={$n_e$ electron number density $[\text{m}^{-3}]$}, grid=major, width = 0.32\textwidth,height=0.33\textwidth, legend pos=south west, title style={at={(0.5,1.1)},anchor=north}, title={t=0.2T}]
					\addplot[-, a1, very thick] table[x={x}, y expr=\thisrow{ne} * 8e16 ] {dat/1dbte_with_E/Ewt_10K_Nx100_Nr127_l2_dt_5e-4_02.csv};
					\addplot[-, a2, very thick] table[x={x}, y expr=\thisrow{ne} * 8e16 ] {dat/1dbte_with_E/Ewt_10K_Nx200_Nr255_l4_dt_2e-4_02.csv};
					\addplot[-, a3, very thick] table[x={x}, y expr=\thisrow{ne} * 8e16 ] {dat/1dbte_with_E/Ewt_10K_Nx200_Nr511_l8_dt_1e-4_02.csv};
					\legend{$r_0$, $r_1$, $r_2$};
				\end{semilogyaxis}
		\end{tikzpicture}}
		\subfloat[\hspace{-0.7in}\label{fig:case1_1_2}]{
			\begin{tikzpicture}
				\begin{semilogyaxis}[xlabel={$\hat{x}$}, ylabel={$n_e$ electron number density $[\text{m}^{-3}]$}, grid=major, width = 0.32\textwidth,height=0.33\textwidth, legend pos=north west, title style={at={(0.5,1.1)},anchor=north}, title={t=0.4T}]
					\addplot[-, a1, very thick] table[x={x}, y expr=\thisrow{ne} * 8e16 ] {dat/1dbte_with_E/Ewt_10K_Nx100_Nr127_l2_dt_5e-4_04.csv};
					\addplot[-, a2, very thick] table[x={x}, y expr=\thisrow{ne} * 8e16 ] {dat/1dbte_with_E/Ewt_10K_Nx200_Nr255_l4_dt_2e-4_04.csv};
					\addplot[-, a3, very thick] table[x={x}, y expr=\thisrow{ne} * 8e16 ] {dat/1dbte_with_E/Ewt_10K_Nx200_Nr511_l8_dt_1e-4_04.csv};
					%\legend{$r_0$, $r_1$, $r_2$};
				\end{semilogyaxis}
		\end{tikzpicture}}
		\subfloat[\hspace{-0.7in}\label{fig:case1_1_3}]{
			\begin{tikzpicture}
				\begin{semilogyaxis}[xlabel={$\hat{x}$}, ylabel={$n_e$ electron number density $[\text{m}^{-3}]$}, grid=major,width = 0.32\textwidth,height=0.33\textwidth, legend pos=north west, title style={at={(0.5,1.1)},anchor=north}, title={t=1.0T}]
					\addplot[-, a1, very thick] table[x={x}, y expr=\thisrow{ne} * 8e16 ] {dat/1dbte_with_E/Ewt_10K_Nx100_Nr127_l2_dt_5e-4_10.csv};
					\addplot[-, a2, very thick] table[x={x}, y expr=\thisrow{ne} * 8e16 ] {dat/1dbte_with_E/Ewt_10K_Nx200_Nr255_l4_dt_2e-4_10.csv};
					\addplot[-, a3, very thick] table[x={x}, y expr=\thisrow{ne} * 8e16 ] {dat/1dbte_with_E/Ewt_10K_Nx200_Nr511_l8_dt_1e-4_10.csv};
					%\legend{$r_0$, $r_1$, $r_2$};
				\end{semilogyaxis}
		\end{tikzpicture}}
		
		\subfloat[\hspace{-0.7in}\label{fig:case1_2_1}]{
			\begin{tikzpicture}
				\begin{semilogyaxis}[xlabel={$\hat{x}$}, ylabel= $n_e$ relative error, grid=major,width = 0.32\textwidth,height=0.4\textwidth, legend pos=south west]
					\addplot[-, a1, very thick] table[x={x}, y expr=\thisrow{ne} ] {dat/1dbte_with_E/Ewt_10K_Nx100_Nr127_l2_dt_5e-4_rel_error_02.csv};
					\addplot[-, a2, very thick] table[x={x}, y expr=\thisrow{ne} ] {dat/1dbte_with_E/Ewt_10K_Nx200_Nr255_l4_dt_2e-4_rel_error_02.csv};
					%\addplot[-, a3, very thick] table[x={x}, y expr=\thisrow{ne} ] {dat/1dbte_with_E/Ewt_10K_Nx200_Nr511_l8_dt_1e-4_rel_error_02.csv};
					\legend{$r_0$ vs. $r_2$, $r_1$ vs. $r_2$};
				\end{semilogyaxis}
		\end{tikzpicture}}
		\subfloat[\hspace{-0.7in}\label{fig:case1_2_2}]{
			\begin{tikzpicture}
				\begin{semilogyaxis}[xlabel={$\hat{x}$}, ylabel= $n_e$ relative error, grid=major,width = 0.32\textwidth,height=0.4\textwidth, legend pos=north west]
					\addplot[-, a1, very thick] table[x={x}, y expr=\thisrow{ne} ] {dat/1dbte_with_E/Ewt_10K_Nx100_Nr127_l2_dt_5e-4_rel_error_04.csv};
					\addplot[-, a2, very thick] table[x={x}, y expr=\thisrow{ne} ] {dat/1dbte_with_E/Ewt_10K_Nx200_Nr255_l4_dt_2e-4_rel_error_04.csv};
					%\addplot[-, a3, very thick] table[x={x}, y expr=\thisrow{ne} ] {dat/1dbte_with_E/Ewt_10K_Nx200_Nr511_l8_dt_1e-4_rel_error_04.csv};
					%\legend{$r_0$, $r_1$, $r_2$};
				\end{semilogyaxis}
		\end{tikzpicture}}
		\subfloat[\hspace{-0.7in}\label{fig:case1_2_3}]{
			\begin{tikzpicture}
				\begin{semilogyaxis}[xlabel={$\hat{x}$}, ylabel= $n_e$ relative error, grid=major,width = 0.32\textwidth,height=0.4\textwidth, legend pos=north west]
					\addplot[-, a1, very thick] table[x={x}, y expr=\thisrow{ne} ] {dat/1dbte_with_E/Ewt_10K_Nx100_Nr127_l2_dt_5e-4_rel_error_10.csv};
					\addplot[-, a2, very thick] table[x={x}, y expr=\thisrow{ne} ] {dat/1dbte_with_E/Ewt_10K_Nx200_Nr255_l4_dt_2e-4_rel_error_10.csv};
					%\addplot[-, a3, very thick] table[x={x}, y expr=\thisrow{ne} ] {dat/1dbte_with_E/Ewt_10K_Nx200_Nr511_l8_dt_1e-4_rel_error_10.csv};
					%\legend{$r_0$, $r_1$, $r_2$};
				\end{semilogyaxis}
		\end{tikzpicture}}
		
		\subfloat[\hspace{-0.7in}\label{fig:case1_3_1}]{
			\begin{tikzpicture}
				\begin{semilogyaxis}[xlabel={$e$-energy [eV]}, ylabel= {$f_0$ [$\text{eV}^{-3/2}$]}, grid=major,width = 0.3\textwidth,height=0.4\textwidth, legend pos=north east, title style={at={(0.5,1.1)},anchor=north}, title={t=0.2T, $\hat{x}$=-0.988}]
					\addplot[-, a1, very thick] table[x={x}, y expr=abs(\thisrow{f0}) ] {dat/1dbte_with_E/Ewt_10K_Nx100_Nr127_l2_dt_5e-4_fl_x-0.988_02.csv};
					\addplot[-, a2, very thick] table[x={x}, y expr=abs(\thisrow{f0}) ] {dat/1dbte_with_E/Ewt_10K_Nx200_Nr255_l4_dt_2e-4_fl_x-0.988_02.csv};
					\addplot[-, a3, very thick] table[x={x}, y expr=abs(\thisrow{f0}) ] {dat/1dbte_with_E/Ewt_10K_Nx200_Nr511_l8_dt_1e-4_fl_x-0.988_02.csv};
					\legend{$r_0$, $r_1$, $r_2$};
				\end{semilogyaxis}
		\end{tikzpicture}}
		\subfloat[\hspace{-0.7in}\label{fig:case1_3_2}]{
			\begin{tikzpicture}
				\begin{semilogyaxis}[xlabel={$e$-energy [eV]}, ylabel= {$f_0$ [$\text{eV}^{-3/2}$]}, grid=major,width = 0.3\textwidth,height=0.4\textwidth, legend pos=north east, title style={at={(0.5,1.1)},anchor=north}, title={t=0.4T, $\hat{x}$=-0.988}]
					\addplot[-, a1, very thick] table[x={x}, y expr=abs(\thisrow{f0}) ] {dat/1dbte_with_E/Ewt_10K_Nx100_Nr127_l2_dt_5e-4_fl_x-0.988_04.csv};
					\addplot[-, a2, very thick] table[x={x}, y expr=abs(\thisrow{f0}) ] {dat/1dbte_with_E/Ewt_10K_Nx200_Nr255_l4_dt_2e-4_fl_x-0.988_04.csv};
					\addplot[-, a3, very thick] table[x={x}, y expr=abs(\thisrow{f0}) ] {dat/1dbte_with_E/Ewt_10K_Nx200_Nr511_l8_dt_1e-4_fl_x-0.988_04.csv};
					%\legend{$r_0$, $r_1$, $r_2$};
				\end{semilogyaxis}
		\end{tikzpicture}}
		\subfloat[\hspace{-0.7in}\label{fig:case1_3_3}]{
			\begin{tikzpicture}
				\begin{semilogyaxis}[xlabel={$e$-energy [eV]}, ylabel= {$f_0$ [$\text{eV}^{-3/2}$]}, grid=major,width = 0.3\textwidth,height=0.4\textwidth, legend pos=north east, title style={at={(0.5,1.1)},anchor=north}, title={t=1.0T, $\hat{x}$=-0.988}, xmax=25]
					\addplot[-, a1, very thick] table[x={x}, y expr=abs(\thisrow{f0}) ] {dat/1dbte_with_E/Ewt_10K_Nx100_Nr127_l2_dt_5e-4_fl_x-0.988_10.csv};
					\addplot[-, a2, very thick] table[x={x}, y expr=abs(\thisrow{f0}) ] {dat/1dbte_with_E/Ewt_10K_Nx200_Nr255_l4_dt_2e-4_fl_x-0.988_10.csv};
					\addplot[-, a3, very thick] table[x={x}, y expr=abs(\thisrow{f0}) ] {dat/1dbte_with_E/Ewt_10K_Nx200_Nr511_l8_dt_1e-4_fl_x-0.988_10.csv};
					%\legend{$r_0$, $r_1$, $r_2$};
				\end{semilogyaxis}
		\end{tikzpicture}}

	\end{center}
	\caption{Electron transport using BTE with specified electric field $E(x,t)=10^4 \sin(2\pi \zeta t)$ with $\zeta=13.56$MHz. \Cref{fig:case1_1_1,fig:case1_1_2,fig:case1_1_3} show the electron number density $n_e(x)=\int_{\vect{v}} f\of{x, \vect{v}}\diff{\vect{x}}$ evolution in time for runs $r_0$, $r_1$, and $r_2$ specified in \Cref{tab:refinement_params}. \Cref{fig:case1_2_1,fig:case1_2_2,fig:case1_2_3} show relative error in $n_e(x)$ with respect to the highest resolution run $r_2$. \Cref{fig:case1_3_1,fig:case1_3_2,fig:case1_3_3} show the electron energy density function at spatial location $\of{2x/L-1}=\hat{x} = -0.988$ for subsequent runs $r_0$, $r_1$ and $r_2$.\label{fig:self_convegence_1d_bte_with_E}}
\end{figure}

We consider the electron BTE with zero incoming flux boundary conditions as specified by~\Cref{eq:hybrid_cnt_b,eq:hybrid_cnt_b_bdy}.
%\begin{subequations}
%    \begin{align}
%        &\partial_t f + v\cos\of{\vtheta} \partial_x f- \frac{q_e E}{m_e} \of{\cos(\vtheta) \partial_v f - \sin(\vtheta) \frac{1}{v} \partial_{\vtheta}f } = C_{en}(f) \text{ for } x \in [0, L]\label{eq:bte_1d_1}, 
%        \\
%        &f(x=0, v, \vtheta \leq \frac{\pi}{2}, \vphi) = 0 \text{ and } f(x=L, v, \vtheta > \frac{\pi}{2}, \vphi) = 0 \label{eq:1d_bte_bc_1}. 
%        %&\vect{E}\of{x,t} = E_0  \sin\of{2\pi f t} \label{eq:bte_with_E_field}. %\of{\frac{2x}{L}-1}^a
%    \end{align}\label{eq:bte_with_E}
%\end{subequations}

We evolve this equation with the initial conditions described by
\begin{align}
    &f(t=0, x, \vect{v}) = \frac{n_e\of{t=0, x}}{{\pi}^{3/2}v_{th}^3} \exp\of{-\of{\frac{\norm{\vect{v}}_2}{v_{th}}}^2} \nonumber \text{ where, } \\ 
    n_e\of{t=0, x} &= 10^{13} + 10^{15} (1- x / L)^2 (x/L)^2 \ [\text{m}^{-3}] \text { ,  } v_{th} = \sqrt{\frac{2k_B T_e}{m_e}} \text{ and } T_e = 0.5 \text{eV}\label{eq:ic_liu},
\end{align} with a specified electric field $E(x,t)$. The electric field causes electron acceleration along the opposite direction, triggering ionization events. Hence, we observe oscillatory growth of electron number density. 

\textit{Convergence}: We use this problem to perform a convergence study. We consider a time-harmonic spatially homogeneous electric field given by $E(x, t) = 10^4 \sin(2\pi \zeta t)$ [V/m] with $\zeta$ = 13.56MHz. \Cref{tab:refinement_params} summarizes the parameters used for the convergence study. \Cref{fig:self_convegence_1d_bte_with_E} shows the convergence of the evolved solution with increasing resolution in the phase space.  \Cref{fig:case1_1_1,fig:case1_1_2,fig:case1_1_3} show the oscillatory growth of the electron number density $n_e(x)$, \Cref{fig:case1_2_1,fig:case1_2_2,fig:case1_2_3} show the relative errors of $n_e(x)$ with respect to the highest resolution run, and \Cref{fig:case1_3_1,fig:case1_3_2,fig:case1_3_3} show the convergence of the electron energy density function at a chosen spatial location. 
\begin{table}[!tbhp]
    \centering
    \begin{tabular}{|c|c|c|c|c|c|}
        \hline
        run id & $N_x$ & $N_r$ & $N_{\vtheta}$ & $N_{l}$ & $\frac{\Delta t}{(1/\zeta)}$ \\
        \hline
        $r_0$ & 100 & 128 & 16 & 2 & 5e-4 \\
        $r_1$ & 200 & 256 & 32 & 4 & 2e-4 \\
        $r_2$ & 200 & 512 & 64 & 8 & 1e-4 \\
        \hline 
    \end{tabular}
    \caption{Refinement parameters used in space, velocity space and time for the convergence study presented in \Cref{fig:self_convegence_1d_bte_with_E}. \label{tab:refinement_params}}
\end{table}

\begin{figure}[!tbhp]
	\begin{center}
		\begin{tikzpicture}
			\begin{semilogyaxis}[xlabel= {$\hat{x} = \frac{2x}{L} - 1$}, ylabel= {number density $[\text{m}^{-3}]$}, grid=major,width = 0.48\textwidth,height=0.46\textwidth, legend pos=south east]
				\addplot[        dotted, a1, very thick]  table[x={x}, y expr=\thisrow{ne_pde} ] {dat/1dbte_with_E_pic_vs_pde/Ewt_10K_pde_idx_0000.csv};
				\addplot[        -     , a2, very thick]  table[x={x}, y expr=\thisrow{ne_pde} ] {dat/1dbte_with_E_pic_vs_pde/Ewt_10K_pde_idx_0010.csv};
				\addplot[        dashed, a3, very thick]  table[x={x}, y expr=\thisrow{ne_pic} ] {dat/1dbte_with_E_pic_vs_pde/Ewt_10K_pic_idx_0010.csv};
				\legend{initial condition, \bte, PIC-DSMC};
			\end{semilogyaxis}
		\end{tikzpicture}
		\begin{tikzpicture}
			\begin{axis}[xlabel = {$\hat{x} = \frac{2x}{L} - 1$}, ylabel= {temperature [eV]}, grid=major,width = 0.48\textwidth,height=0.46\textwidth, legend pos=south east]
				\addplot[        dotted, a1, very thick]  table[x={x}, y expr=\thisrow{Te_pde} ] {dat/1dbte_with_E_pic_vs_pde/Ewt_10K_pde_idx_0000.csv};
				\addplot[        -     , a2, very thick]  table[x={x}, y expr=\thisrow{Te_pde} ] {dat/1dbte_with_E_pic_vs_pde/Ewt_10K_pde_idx_0010.csv};
				\addplot[        dashed, a3, very thick]  table[x={x}, y expr=\thisrow{Te_pic} ] {dat/1dbte_with_E_pic_vs_pde/Ewt_10K_pic_idx_0010.csv};
				%\legend{initial condition, Eulerian, PIC-DSMC};
			\end{axis}
		\end{tikzpicture}
	\end{center}
	\caption{We consider the electron BTE given in \Cref{eq:hybrid_cnt_b,eq:hybrid_cnt_b_bdy} with electric field $E(x,t)=10^4 \sin(2\pi \zeta t)$ with $\zeta=13.56$MHz. The figure shows solutions after a one period 1/f. The blue line is~\bte~and the broken green line is PIC-DSMC. \label{fig:case1_pic_vs_pde}}
\end{figure}
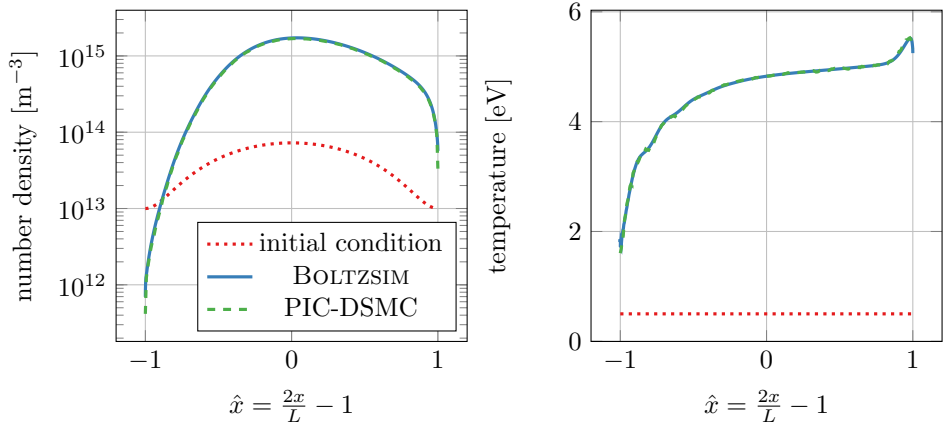

We solve the electron BTE given in \Cref{eq:hybrid_cnt_b,eq:hybrid_cnt_b_bdy} with our \bte~and PIC-DSMC codes and we report the results in \Cref{fig:case1_pic_vs_pde}. The results show excellent agreement between the two solvers. 
%In the above, the solutions computed using Eulerian and PIC-DSMC approaches show excellent agreement. 

\subsection{Performance evaluation}
\label{subsec:peformance_glowd}
We summarize performance evaluation results for the fluid and \bte~solvers for RF-GDPs. We conducted our experiments on Lonestar 6 at the Texas Advanced Computing Center (TACC). Each node on Lonestar 6 has two AMD EPYC 7763 CPUs and three NVIDIA A100 GPUs. 
\begin{table}[!tbhp]
	\centering
	\begin{tabular}{||c|c|c||}
		\hline
		$N_x$ & $\Delta t / (1/\zeta) $  & timestep (s) \\
		\hline
		100	  & 1.00E-03        & 9.0785E-03 \\
		200	  & 1.00E-03        & 2.9685E-02 \\
		400	  & 1.00E-03        & 1.7288E-01 \\
		\hline
	\end{tabular}
	\caption{The fluid solver runtime for a single implicit timestep with increasing spatial resolution. The above results are conducted using a single core of AMD EPYC 7763 CPU \label{tab:performance_fluid}.}
\end{table}
The presented fluid and \bte~performance results are conducted on a single AMD EPYC 7763 CPU core and a single NVIDIA A100 GPU. \Cref{tab:performance_fluid} summarizes the overall implicit timestep cost for the fluid solver. We observe $\mathcal{O}(N_x^3)$ growth of the timestep cost where $N_x$ denotes the number of Chebyshev collocation points in space. This is due to the direct Jacobian factorization used in the Newton iteration. 

Performance evaluation results of the \bte~solver is presented in \Cref{tab:performance_hybrid}. 
\begin{table}[!tbhp]
	\centering
	\resizebox{\textwidth}{!}{
		\begin{tabular}{||c|c|c|c|c|c|c|c|c||}
			\hline
			$N_r$ & $N_\vtheta$ & $N_l$ & $N_x$ & $\Delta t_S \times f$ & $\Delta t_F \times f$ & semi-implicit (s) & fully-implicit (s) & speedup\\
			\hline
			64  & 16 & 3 & 100 & 5E-05 & 2E-04 & 3.53E-02 &  6.52E-02 & 2.1x\\
			128 & 16 & 3 & 100 & 5E-05 & 2E-04 & 3.61E-02 &  6.55E-02 & 2.2x\\
			256 & 16 & 3 & 100 & 5E-05 & 2E-04 & 3.71E-02 &  6.68E-02 & 2.2x\\
			\hline
			\hline
			64  & 16 & 3 & 100 & 5E-05 & 2E-04 & 3.53E-02 &  6.52E-02 & 2.1x\\
			128 & 16 & 3 & 200 & 5E-05 & 2E-04 & 3.93E-02 &  2.52E-01 & 0.6x \\
			256 & 16 & 3 & 400 & 5E-05 & 2E-04 & 6.37E-02 &  --& --\\
			\hline
	\end{tabular}}
	\caption{Comparison of the overall timestep cost for the \bte~solver using semi-implicit and fully-implicit schemes. Although the fully implicit solver is more expensive per timestep, it is faster overall because it requires fewer timesteps. The last column shows the overall runtime (semi-implicit / fully-implicit). The fully-implicit scheme is $\sim2\times$ faster than the semi-implicit scheme for coarser spatial grids $N_x$=100. For $N_x\geq 200$ the fully-implicit solve becomes too expensive.
		%The efficiency $S$ of the fully-implicit scheme reduces with increasing spatial resolution $N_x \geq $ 200
		\label{tab:performance_hybrid}}
\end{table}
Here, $N_x$, $N_r$, $N_\vtheta$, and $N_l$ denote the number of Chebyshev collocation points in the position space, B-spline basis functions in speed, discrete ordinate points, and spherical basis functions are used in $\vtheta$. For coarser spatial grids, we observe that the fully-implicit scheme is $\sim 2\times$ faster than the semi-implicit scheme. With increased spatial resolution, the fully-implicit scheme becomes more expensive than the semi-implicit scheme. This is primarily due to the increased $T_{\vect{x}\text{-rhs}}$ cost. Recall that the  complexities of fully-implicit and semi-implicit schemes are given by $2k(T_{\vect{x}\text{-rhs}} + T_{\vect{v}\text{-rhs}})$ and $(T_{\vect{x}\text{-rhs}} + 2k T_{\vect{v}\text{-rhs}})$ where $k$ denotes the number of iterations for convergence. Hence, the increased $T_{\vect{x}\text{-rhs}}$ reduces the overall efficiency of the fully-implicit scheme. Of course both methods are orders of magnitude faster than an explicit method. We use the coarse grid fully-implicit scheme to march over the transient and reach the time-periodic state. Once time-periodicity is reached, we use iterative refinement with semi-implicit time integration to resolve the finer features of the time-periodic state.

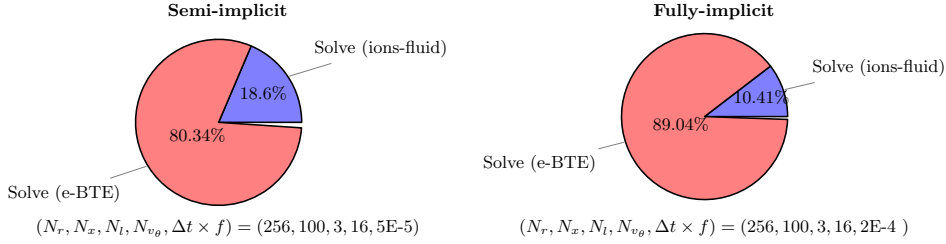
\begin{figure}[!tbhp]
	\centering
	\small
	\resizebox{\textwidth}{!}{
	\begin{tabular}{cc}
		\textbf{Semi-implicit} & \textbf{Fully-implicit}\\
		\begin{tikzpicture}
			\pie[radius=1.5, text=pin,color={blue!50,red!50}]{18.6/\text{Solve (ions-fluid)}, 80.34/\text{Solve (e-BTE)}};
		\end{tikzpicture} & \begin{tikzpicture}
			\pie[radius=1.5,text=pin, color={blue!50,red!50}]{10.41/\text{Solve (ions-fluid)}, 89.04/\text{Solve (e-BTE)}};
		\end{tikzpicture}\\
	$(N_r,N_x,N_l,N_{\vtheta}, \Delta t \times f)=(256, 100, 3, 16, \text{5E-5})$ & $(N_r,N_x,N_l,N_{\vtheta}, \Delta t \times f)=(256, 100, 3, 16, \text{2E-4 })$\\
	\end{tabular}}
\caption{The overall runtime cost breakdown for a single RF-GDP timestep using the hybrid modeling approach. The left and right plots show the cost breakdown using semi-implicit and fully-implicit schemes for evolving the electron BTE. In both cases, $\vect{E}$-field solve cost is negligible compared to others, hence omitted in the plot. \label{fig:cost_breakdown}}
\end{figure}

\Cref{fig:cost_breakdown} shows the overall cost breakdown for a single timestep for evolving the hybrid formulation of RF-GDPs with semi-implicit and fully-implicit time integration schemes. We compute these percentages based on each method's overall runtime for a single complete timestep to evolve ions and electrons. Even though the relative cost of the BTE update is 10\% higher in the fully-implicit scheme, compared to the semi-implicit approach, it allows larger timestep sizes that respect timescales of ions. 

\subsection{Glow discharge}
\label{subsec:results_glowd}
\begin{figure}[!tbhp]
	\centering
	\begin{tikzpicture}
		\tikzstyle{d}=[rectangle,
		rounded corners=8pt,
		very thick,
		text centered,
		minimum size=1cm,
		draw=black!100,
		fill=green!10];
		\node[text width=2cm, d](A) at (0,0) {$n_0$, $T_0$ with $\{E_0, ..., E_{n}\}$};
		\node[text width=5cm, d](B) at (4.5,0) {Solve: $\partial_t f - \frac{q_eE}{m_e} \nabla_{\vect{v}} f = C_{en}(f)$};
		\node[d](C) at (9.5, 0)
		{		
				\begin{tabular}{|c|c|c|c|}
					\hline
					$T_e$ & $k_i$ & $\mu_e$ & $D_e$\\
					\hline
					$T_e^0$ & $k_i^0$ & $\mu_e^{0}$ & $D_e^{0}$ \\
					\vdots & \vdots& \vdots & \vdots\\
					$T_e^n$ & $k_i^n$ & $\mu_e^{n}$ & $D_e^{n}$ \\
					\hline
				\end{tabular}
		};
	\draw[->, line width=0.5mm] (A) -- (B);
	\draw[->, line width=0.5mm] (B) -- (C);
	\end{tikzpicture}
\caption{An illustrative diagram that describes the process of computing tabulated kinetic coefficients for electrons to be used in the fluid model (Equation~\ref{eq:fluid_model_eqs}). For a fixed $n_0$, $T_0$, we compute the steady-state EDF solutions for the spatially homogeneous BTEs for a given electric field values i.e., $\{E_0,...,E_n\}$. For each $E_j$, we compute kinetic coefficients \{$k_i^j$, $\mu_e^j$, $D_e^j$\} and corresponding temperature $T_e^j$. Then for example, for $\mu_e$, we build an interpolant $\mu_e(T_e)$ using the $\{\mu_e^j, T_e^j\}$ pairs. 
%This will result in electron kinetic coefficients (i.e., $k_i$, $\mu_e$, and $D_e$) for the corresponding steady-state electron temperature $T_e$ values (i.e., $\{T_e^{0},...,T_e^{n}\}$). We use the above to build electron temperature-based tabulated kinetic coefficients that are used in the fluid approximation.
\label{fig:tab_kinetics_process} 
}
\end{figure}
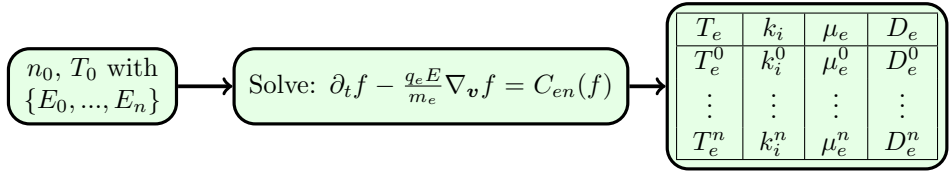

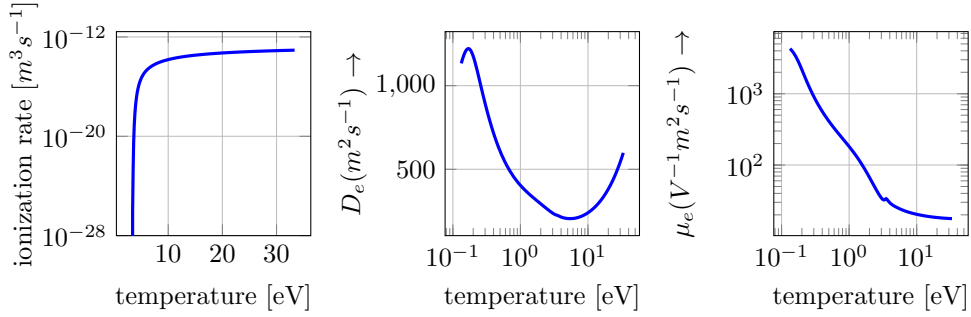
\begin{figure}[!tbhp]
	\centering
	\begin{tikzpicture}
		\begin{semilogyaxis}[xlabel={temperature [eV]}, ylabel={ionization rate $[m^3s^{-1}]$}, grid=major, width = 0.32\textwidth, height=0.33\textwidth, ymin=1e-28]
			\addplot[blue, very thick] table[x expr = \thisrow{Te(K)}/11604.518, y expr=1e-100 + \thisrow{kf(m3/s)}, col sep=comma]{dat/Ar_1Torr_300K/Ionization.300K.txt};
		\end{semilogyaxis}
	\end{tikzpicture}
	\begin{tikzpicture}
		\begin{semilogxaxis}[xlabel={temperature [eV]}, ylabel= $D_e (m^{2}s^{-1})$ $\rightarrow$, grid=major, width = 0.32\textwidth, height=0.33\textwidth]
			\addplot[blue, very thick] table[x expr = \thisrow{Te(K)}/11604.518, y expr = \thisrow{D*N(1/m-s)}/3.22e22, col sep=comma]{dat/Ar_1Torr_300K/Diffusivity.300K.txt};
		\end{semilogxaxis}
	\end{tikzpicture}
	\begin{tikzpicture}
		\begin{loglogaxis}[xlabel={temperature [eV]}, ylabel= $\mu_e (V^{-1} m^{2} s^{-1})$ $\rightarrow$, grid=major,width = 0.32\textwidth, height=0.33\textwidth]
			\addplot[blue, very thick] table[x expr = \thisrow{Te(K)}/11604.518, y expr =\thisrow{Mu*N(1/V-m-s)} / 3.22e22, col sep=comma]{dat/Ar_1Torr_300K/Mobility.300K.txt};
		\end{loglogaxis}
	\end{tikzpicture}
	\caption{Tabulated electron kinetic coefficients based on the steady-state solutions of the 0D3V electron BTE. \label{fig:bte_0d_kinetics}}
\end{figure}

For modeling RF-GDPs, we present a comparative study of the fluid and the hybrid approaches, which are given by~\Cref{eq:fluid_model_eqs,eq:glow_hybrid}. The key difference between the fluid and hybrid modeling approaches is how the electron transport is being modeled. The fluid approach transports electrons and ions using fluid equations. In contrast, in the hybrid model, electron transport is kinetic by solving the BTE, while we use the standard fluid transport for ions. We consider four cases with varying gas pressure values ranging from 0.1 Torr to 2 Torr while keeping other parameters unchanged. Since $n_0 \sim \frac{p_0}{k_B T}$, lower pressure values result in lower background number densities. This causes the plasma to be less collisional. Our primary goal is to identify the modeling discrepancies between hybrid and fluid-based electron transport under different pressure regimes. The simulation parameters used for the study are summarized in \Cref{t:model_parameters}. 
\begin{table}[!tbhp]
	\centering
	\resizebox{\textwidth}{!}{
	\begin{tabular}{|c|c|c|c|}
		\hline 
		\hline
		Parameter & Symbol &  Value & Reference \\
		\hline
		\small Gas pressure     & $p_0$ & 0.1Torr, 0.5Torr, 1Torr, 2Torr  & --\\
		\small Gas temperature  & $T_0$ & 300 K & --\\
		\small Neutral density  & $n_0$ & (0.322, 1.61, 3.22, 6.44)$\times 10^{22} \text{m}^{-3}$  & ideal gas law\\
		\small Electrode gap    & $L$   & 2.54 $\times 10^{-2}$ $\text{m}$ & \cite{liu2014numerical}\\
		\small Peak voltage     & $V_0$ & 100 V & \cite{liu2014numerical} \\
		\small Oscillation frequency & $f$ & 13.56 MHz & \cite{lymberopoulos1993fluid}\\
		\small Electron diffusion  & $D_e$  & \bolsig~ (0D-BTE) & \cite{hagelaar2005solving, hagelaar2015coulomb}\\
		\small Ion diffusion   & $D_i$ & $n_0 D_i = 2.07 \times 10^{20} \text{m}^{-1}\text{s}^{-1}$ & \cite{lymberopoulos1993fluid} \\
		\small Electron mobility  & $\mu_e$ & \bolsig~ (0D-BTE) & \cite{hagelaar2005solving, hagelaar2015coulomb}\\
		\small Ion mobility    & $\mu_i$ & $n_0 \mu_i = 4.65 \times 10^{21} \text{V}^{-1}\text{m}^{-1}\text{s}^{-1}$ & \cite{lymberopoulos1993fluid}\\
		\small $e + Ar \rightarrow 2e + Ar^{+}$ & $k_i$ & \bolsig~ (0D-BTE) & \cite{hagelaar2005solving, hagelaar2015coulomb}\\
		\hline
		\hline
	\end{tabular}}
	\caption{Modeling parameters used in the presented work. \label{t:model_parameters}}
\end{table}
The fluid approximation-based species transport relies on a closure model for kinetic coefficients. The simplest closure model would be the Maxwellian EDF assumption and derive temperature-based electron kinetic coefficients~\cite{liu2014numerical}. The above is not ideal due to the electric field effects in RF-GDPs and other LTPs because of strong deviation from Maxwellian and non-local effects that can occurs at low pressures. In our fluid approximation, we allow non-Maxwellian treatment of kinetic coefficients by solving spatially homogeneous electron BTE. We use the state-of-the-art \bolsig~framework~\cite{hagelaar2005solving} to compute steady-state EDFs for varying $\norm{\vect{E}}$ values with corresponding neutral density $n_0$ and gas temperature $T_0$. With the computed steady-state EDFs, we build electron temperature-based look-up tables for required electron kinetics. The above process is illustrated in \Cref{fig:tab_kinetics_process}. \Cref{fig:bte_0d_kinetics} shows the tabulated kinetic coefficients plotted against the electron temperature $T_e$. Following~\cite{liu2014numerical} we use constant $D_i$, $\mu_i$ for ions. 
%For heavy species transport, we use constant kinetic coefficients similar to ~\cite{liu2014numerical}.

For RF-GDPs, the time-periodic steady-state solution is determined by the background gas temperature $T_0$, pressure $p_0$, the electrode gap $L$, and the peak driving voltage $V_0$. As mentioned, we consider the $p_0$ values listed in \Cref{t:model_parameters} while keeping all other parameters fixed. Changing $p_0$ changes the neutral density $n_0$, which is coupled to the problem through electron-heavy binary collisions and heavy species kinetic coefficients. 

Both hybrid and fluid solvers were time-marched until the time-periodic state is achieved. We compute the time-periodic solution from the fluid solver with initial conditions given by~\cite{lymberopoulos1993fluid}. This solution is then used as an initial condition for the hybrid solver, where the EDFs are initialized as Maxwellian distributions defined by the temperature computed by the fluid solver. %The primary difference between the fluid and the hybrid approaches is the electron transport. 

%We evolve the fluid and hybrid RF-GDP models for the parameters summarized in \Cref{t:model_parameters} until the time-periodic steady-state is achieved. Note that, for the fluid approximation, we allow non-Maxwellian treatment of EDFs through tabulated kinetic coefficients computed by solving spatially homogeneous BTE for local plasma conditions. First, we compute the time-periodic steady-state solution from the fluid solver where the initial conditions are similar to~\cite{lymberopoulos1993fluid}. The time-periodic steady-state solution predicted by the fluid approximation is used as an initial condition for the hybrid solver, where EDFs are initialized for Maxwellian distribution with the local temperature. The key difference between the fluid and hybrid modeling approaches is how the electron transport is being modeled. The fluid approach transports electrons and ions using fluid equations. In contrast, in the hybrid model, electron transport is entirely handled by the BTE, and the fluid approximation handles the transport of ions. 

\begin{figure}[!tbhp]
	\includegraphics[width=\textwidth]{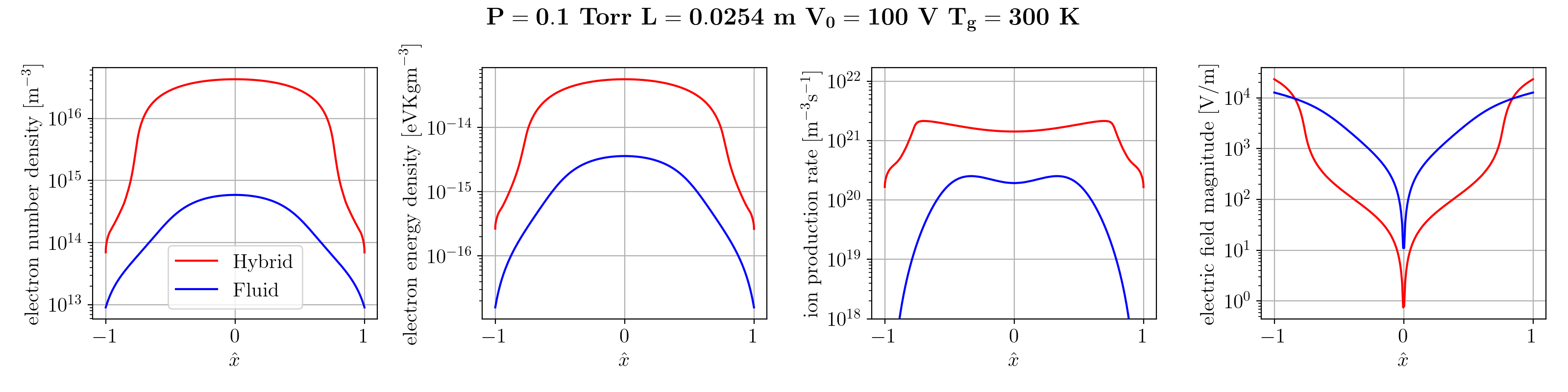} \\
	\includegraphics[width=\textwidth]{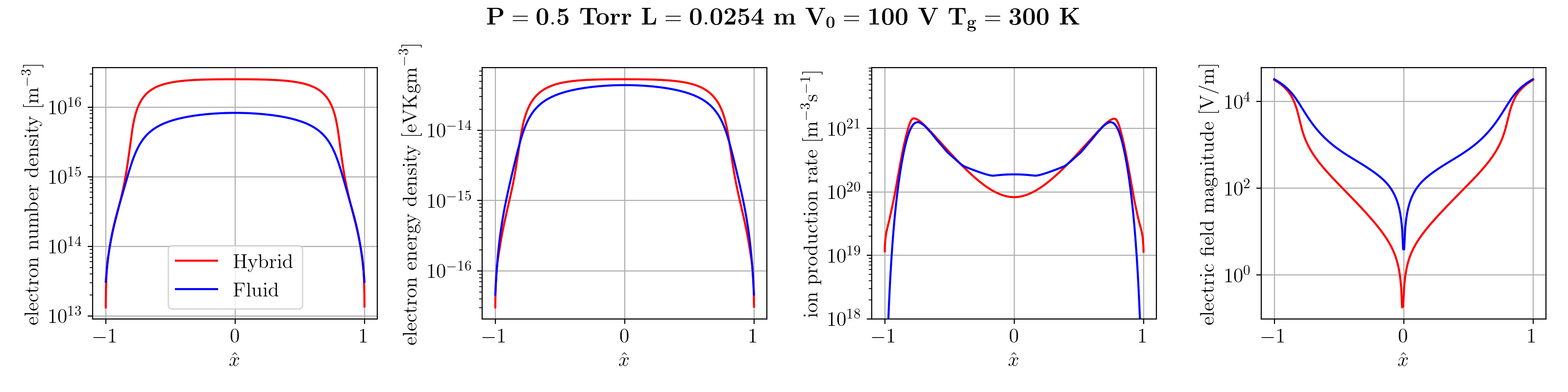} \\
	\includegraphics[width=\textwidth]{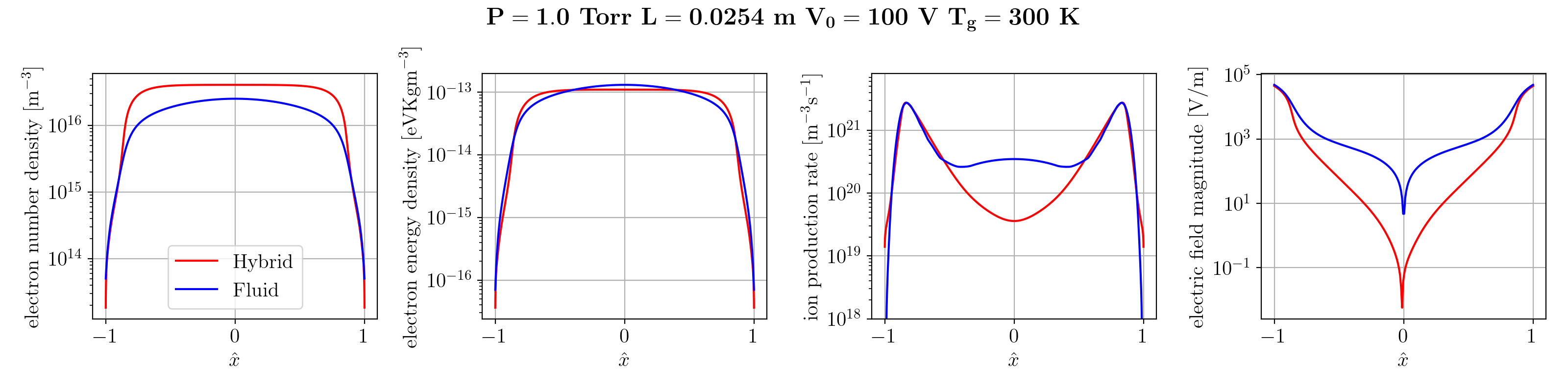} \\
	\includegraphics[width=\textwidth]{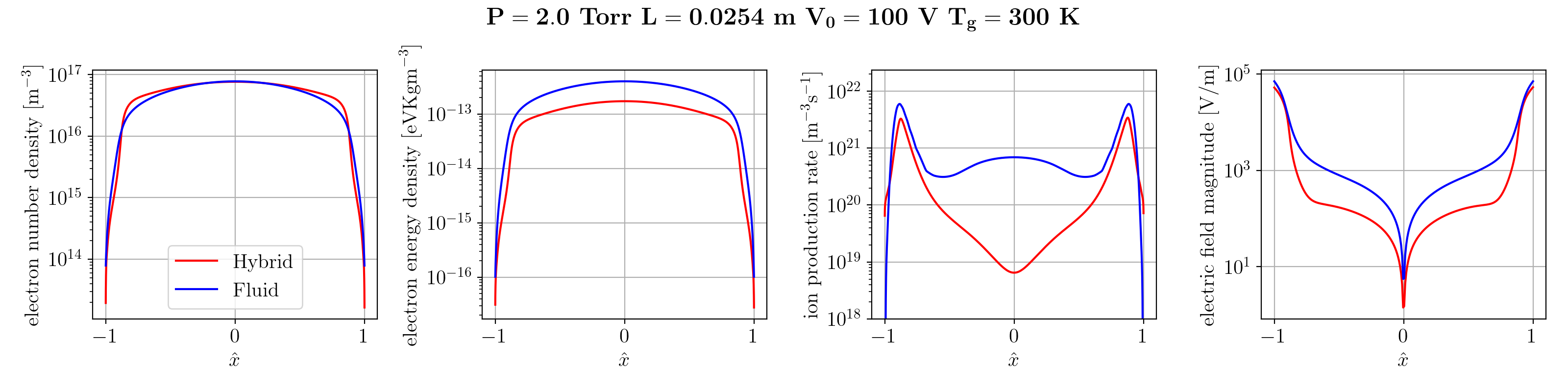} 
	\caption{Cycle-averaged time-periodic steady-state profiles computed by the hybrid and the fluid approximation of RF-GDPs under varying pressure values from 0.1 Torr to 2 Torr. \label{fig:ca_profiles_fluid_vs_hybrid} }
\end{figure}
\begin{table}[!tbhp]
	\centering
	\resizebox{\textwidth}{!}{
		\begin{tabular}{|c|m{2.3cm}|m{2.3cm}|m{2.3cm}|m{2.3cm}|m{2.3cm}|m{2.3cm}|}
			\hline 
			&  \multicolumn{3}{c|}{\textbf{Hybrid - cycle averaged peak}}  &  \multicolumn{3}{c|}{\textbf{Fluid - cycle averaged peak}} \\
			\hline
			$\bf{p_0}\ [Torr]$ & $\bf{n_e}\ [m^{-3}]$ & $\bf{\varepsilon_e}\ [eV kg m^{-3}]$ & $\bf{|E|}\ [V/m]$ & $\bf{n_e}\ [m^{-3}]$ & $\bf{\varepsilon_e}\ [eV kg m^{-3}]$ & $\bf{|E|}\ [V/m]$\\
			\hline
			0.1 & 4.27E+16 & 5.65E-14 & 2.32E+04 & 5.87E+14 & 3.660E-15 & 1.27E+04 \\
			\hline
			0.5 & 2.51E+16 & 5.33E-14 & 3.15E+04 & 8.18E+15 & 4.35E-14 & 3.23E+04 \\
			\hline
			1.0 & 4.06E+16 & 1.10E-13 & 4.44E+04 & 2.51E+16 & 1.31E-13 & 4.83E+04\\
			\hline
			2.0 & 7.67E+16 & 1.72E-13 & 5.32E+04 & 7.79E+16 & 3.98E-13 & 7.06E+04 \\
			\hline
	\end{tabular}}
	\caption{A summary of the peak values for the cycle-averaged time-periodic steady-state profiles computed by fluid and hybrid modeling approaches. \label{tab:fluid_vs_hybrid}}
\end{table}

The cycle-averaged profiles are shown in \Cref{fig:ca_profiles_fluid_vs_hybrid}. \Cref{tab:fluid_vs_hybrid} summarizes the peak values for the cycle-averaged time-periodic solutions computed by fluid and hybrid models. For the cycle-averaged profiles, the largest discrepancy between the fluid and the hybrid model is reported in the lowest pressure case, $p_0$ = 0.1 Torr. Specifically, the hybrid computed peak cycle-averaged steady-state electron number density is 80x, 3x, and 1.6x higher compared to the equivalent fluid runs for 0.1 Torr, 0.5 Torr, and 1 Torr cases. For the highest pressure case presented, $p_0$ = 2 Torr, the cycle-averaged electron number density profiles agree quite well but there are still discrepancies, for example the peak electric field. Below we list the reasons that contribute to this discrepancies. 
%Due to the non-linearity, and non-local coupling of modeling equations (i.e., both fluid and hybrid approaches) is it not possible to idle a single cause to explain the differences we observe between models, but we can list the following reasons that cumulatively contribute to the observed differences. 

\begin{figure}[!tbhp]
	\includegraphics[width=0.24\textwidth]{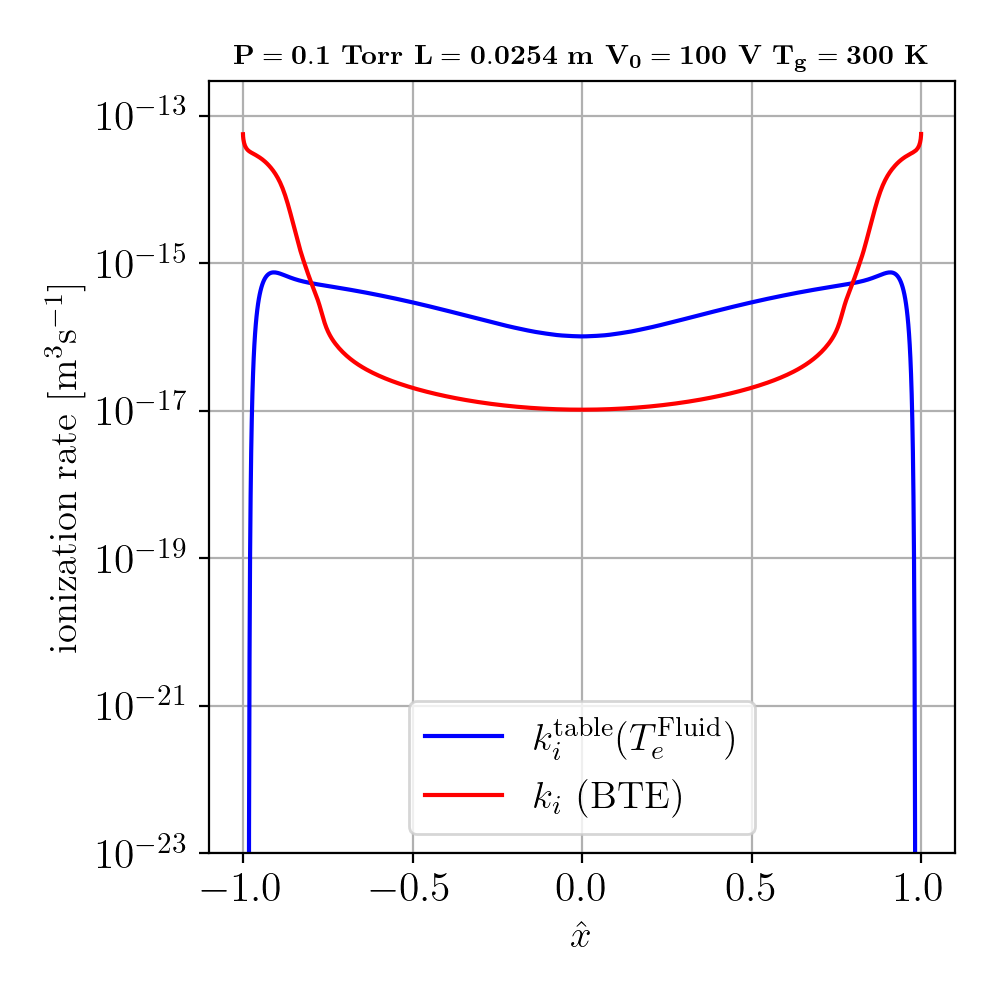} 
	\includegraphics[width=0.24\textwidth]{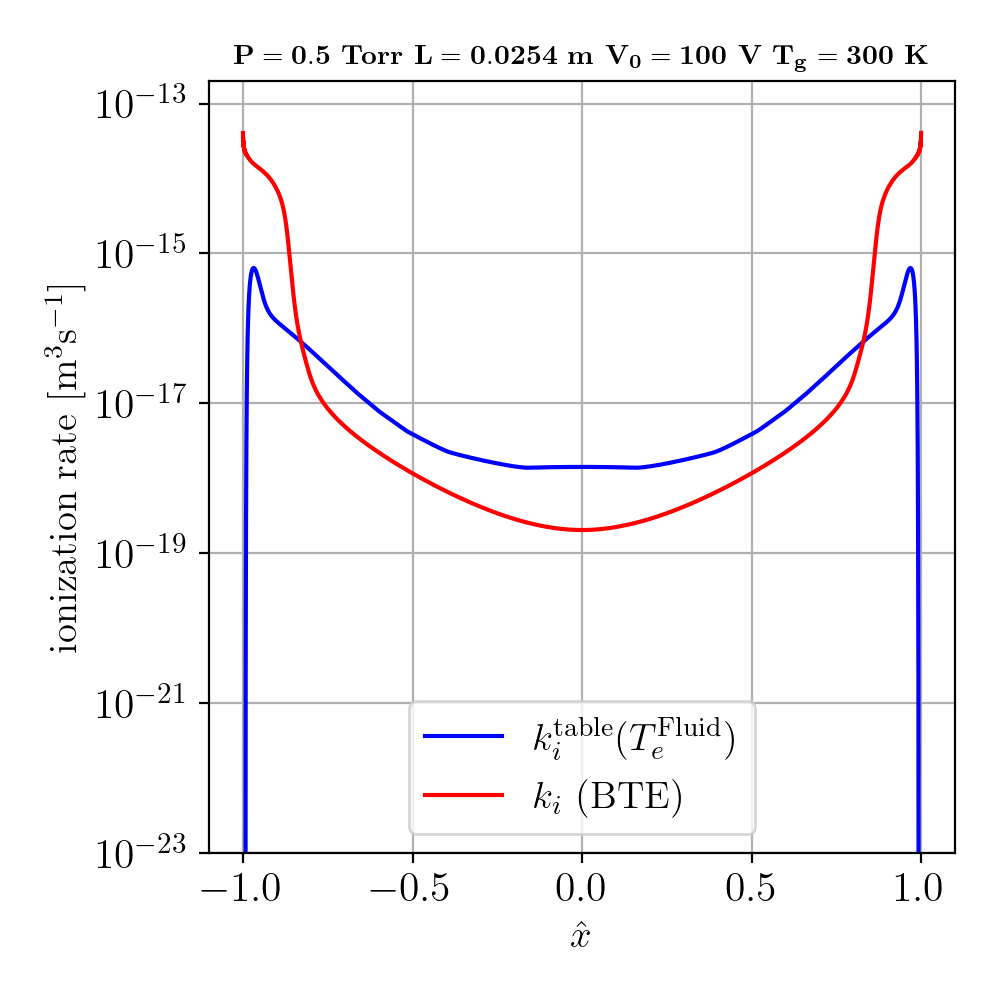} 
	\includegraphics[width=0.24\textwidth]{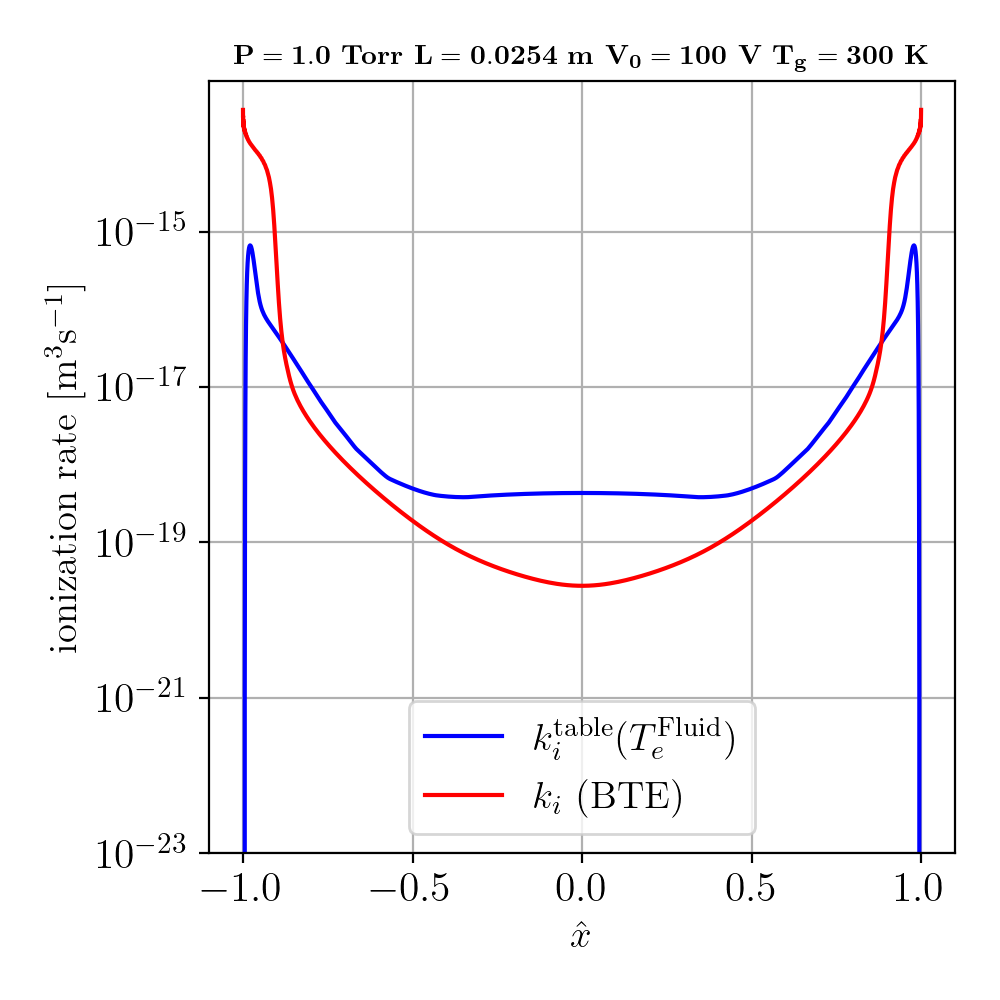}   
	\includegraphics[width=0.24\textwidth]{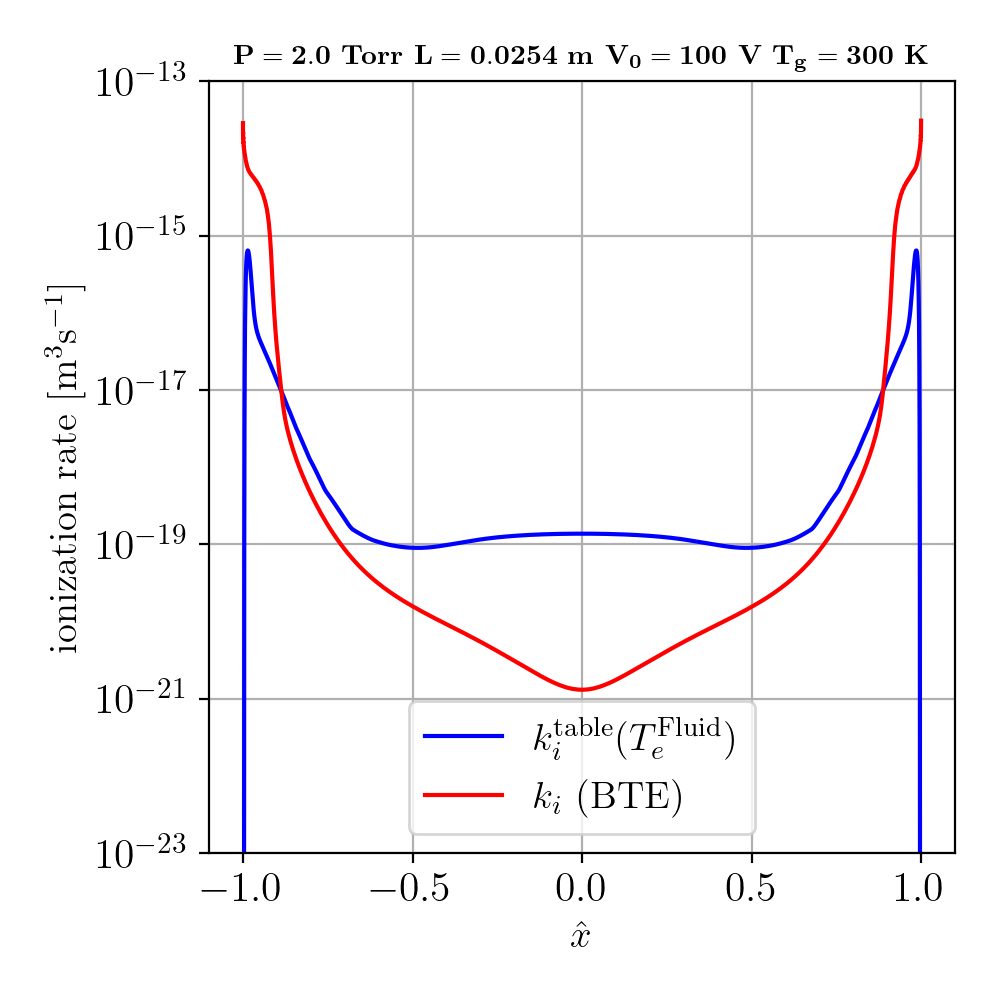}   
	\caption{Discrepancies in the ionization rate coefficient computed from the tabulated data generated form steady-state spatially homogeneous BTE and spatially coupled BTE in the hybrid modeling approach. \label{fig:ki_fluid_vs_hybrid}}
\end{figure}

\begin{figure}[!tbhp]
	\centering
	\includegraphics[width=\textwidth]{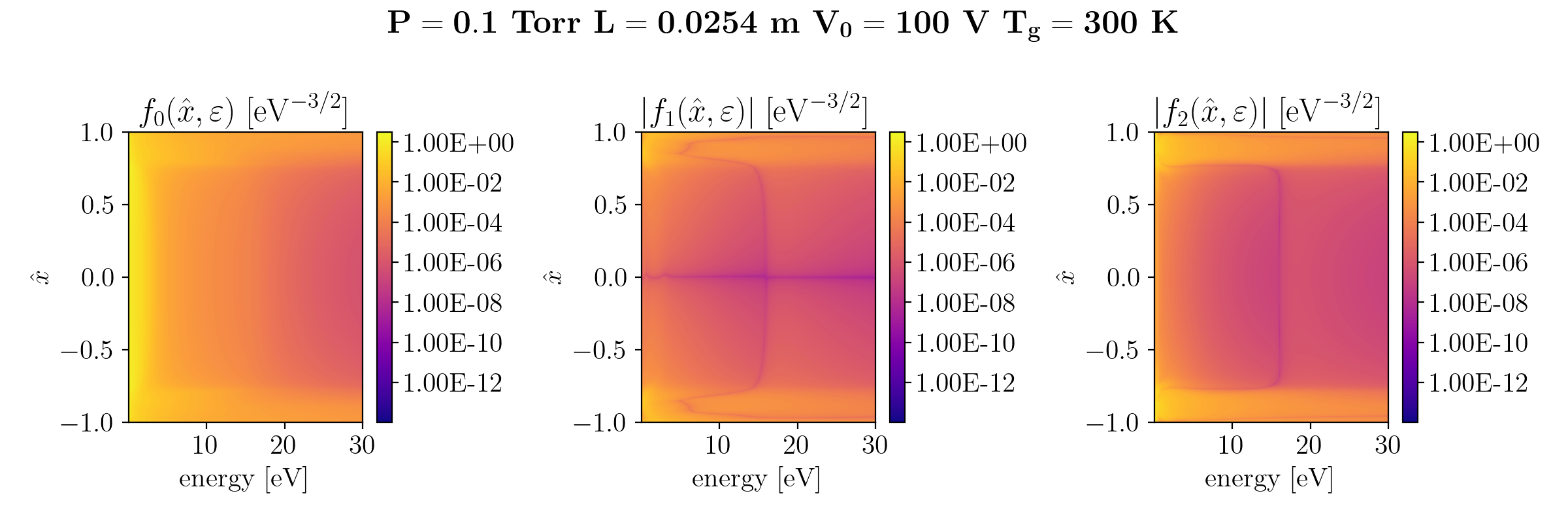}
	\includegraphics[width=\textwidth]{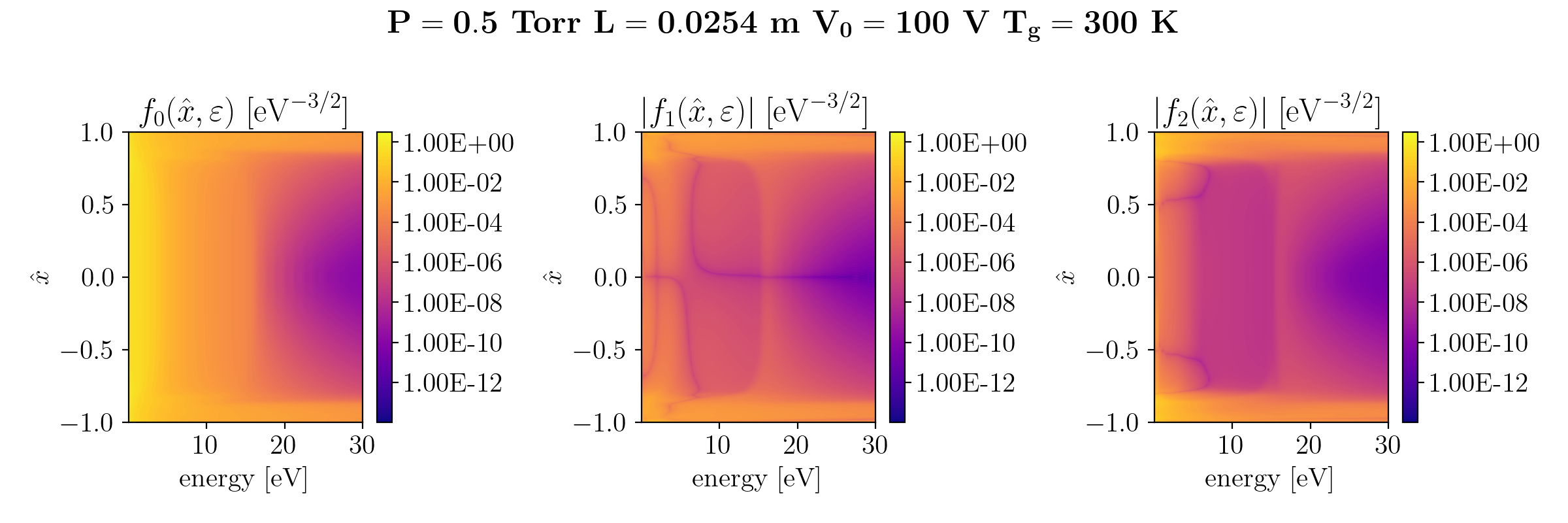}
	\includegraphics[width=\textwidth]{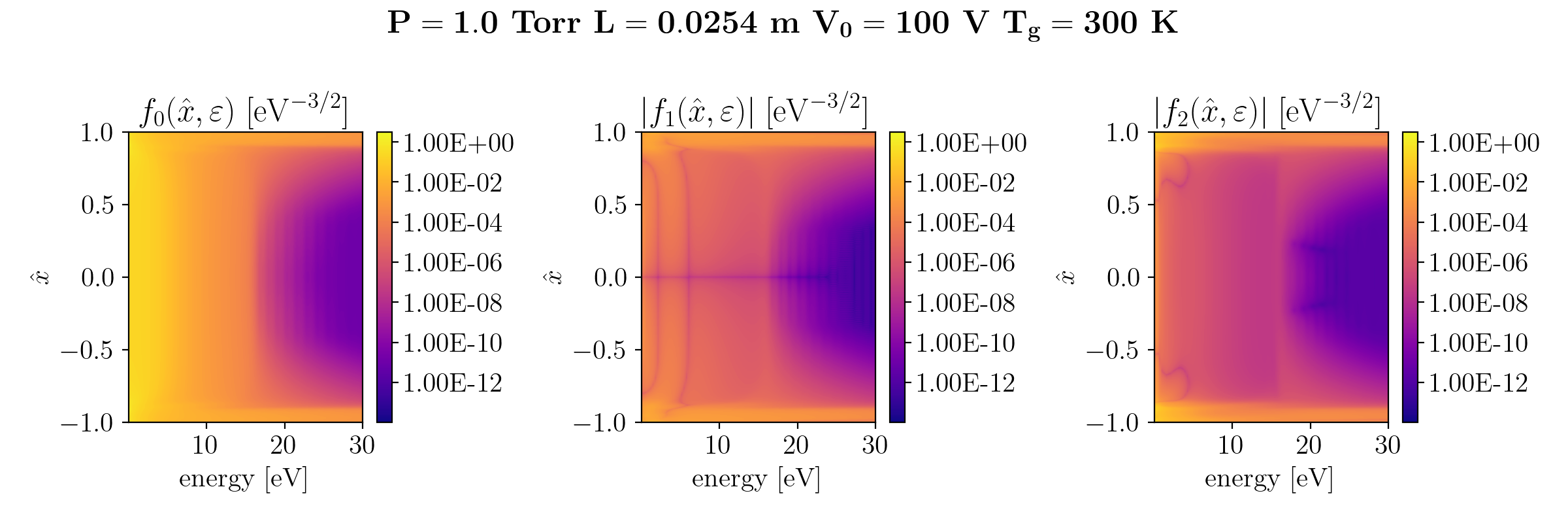}
	\includegraphics[width=\textwidth]{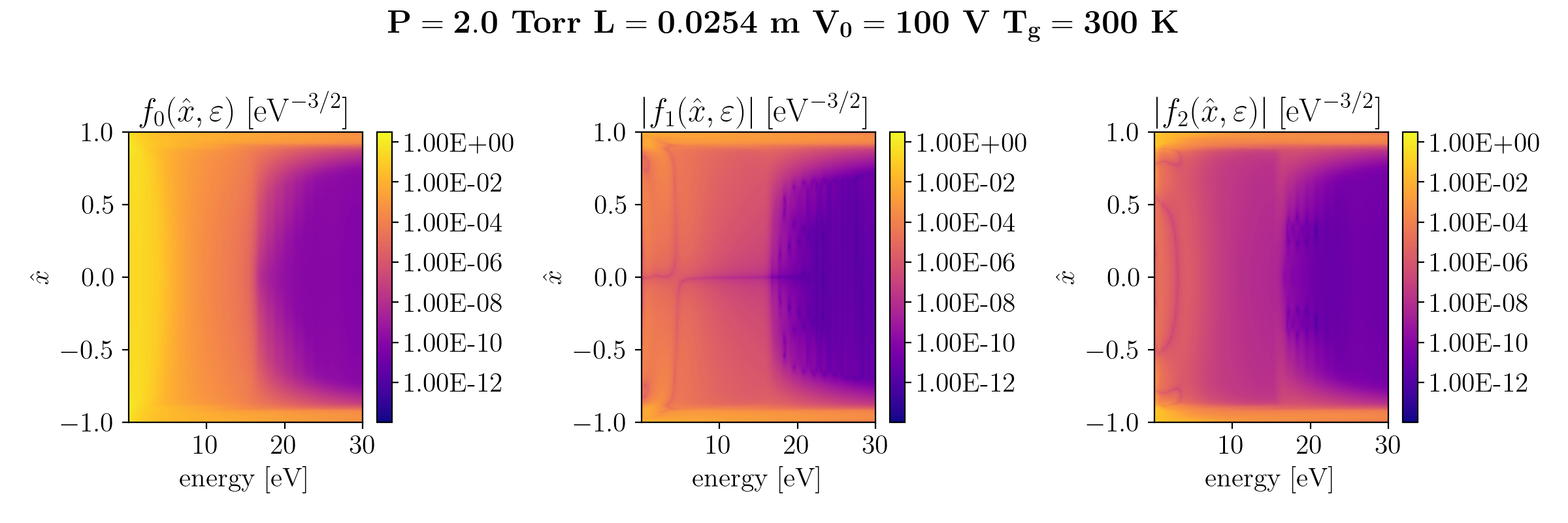}
	\caption{Cycle-averaged time-periodic steady-state EDF solutions in RF-GDPs for the different pressure values computed using the hybrid modeling approach. \label{fig:eedf_hybrid}}
\end{figure}

\textbf{Tabulated rate coefficients}: The fluid approximation of the RF-GDPs relies on a closure model for the electron kinetic coefficients. In the fluid solve, electron kinetics are computed using tabulated results using the 0D3V BTE while for hybrid model, kinetics coefficients are computed by the 1D3V BTE. 
%For the presented fluid model, the tabulated kinetic coefficients are constructed by solving a series of spatially homogeneous electron BTEs for steady-state EDFs with varying electric field magnitudes. The above data is used to construct one-dimensional tabulated kinetic coefficients, which are queried by the corresponding steady-state electron temperature. \Cref{fig:ki_fluid_vs_hybrid} shows the ionization rate coefficient computed by the tabulated data with the fluid approximated electron temperature ($T_e^\mathrm{Fluid}$) and using the EDF computed by solving one-dimensional electron BTE in the hybrid modeling approach.
%We observe that, at the sheath, the tabulated rate coefficient computed by using $T_e^\mathrm{Hybrid}$ matches closely with the rate coefficient computed by solving one-dimensional BTE. The above is due to the large electric field presence at the sheath region reduces the relative effect of the spatial coupling terms in the BTE. At the core of the discharge, we observe that the tabulated rate coefficients cannot produce ions with the electron temperature reported by the hybrid model. The above is primarily due to the fact that the spatial coupling effects reduce tail depletion due to collisions, enabling to trigger ionization reactions with a lower core temperature. 
\Cref{fig:eedf_hybrid} shows the radial components of the cycle-averaged steady-state EDFs computed by the hybrid solver. With increasing background gas pressure, the EDFs at the center of the discharge show rapidly depleted tails at the ionization energy threshold value 15.76 eV. Specifically, for the 0.1 Torr case, tails are more prominent due to reduced collisionality. Similarly, the anisotropic correction modes are dominant compared to runs with higher background gas pressure. Furthermore, tabulated data-based kinetic coefficients
are inaccurate for low-pressure cases because the 0D3V BTE misses spatial coupling effects. These effects become more pronounced with decreasing pressure. 

\begin{figure}[!tbhp]
	\centering
	%\includegraphics[width=\textwidth]{fig/0.1Torr300K_100V_Ar_3sp2r_cycle_uz.png} \\
	%\includegraphics[width=\textwidth]{fig/0.5Torr300K_100V_Ar_3sp2r_cycle_uz.png} \\
	%\includegraphics[width=\textwidth]{fig/1Torr300K_100V_Ar_3sp2r_cycle_uz.png}   \\
	%\includegraphics[width=\textwidth]{fig/2Torr300K_100V_Ar_3sp2r_cycle_uz.png}   
	%u_z \caption{Comparison of the electron drift velocity magnitude ($||\vect{u}_e||_2$) computed based on the hybrid and the fluid approximation modeling approaches. The third column shows the relative error between the two approaches taking the hybrid model as the true solution. \label{fig:drift_velocity_hybrid_fluid}}
	
	\includegraphics[width=0.46\textwidth]{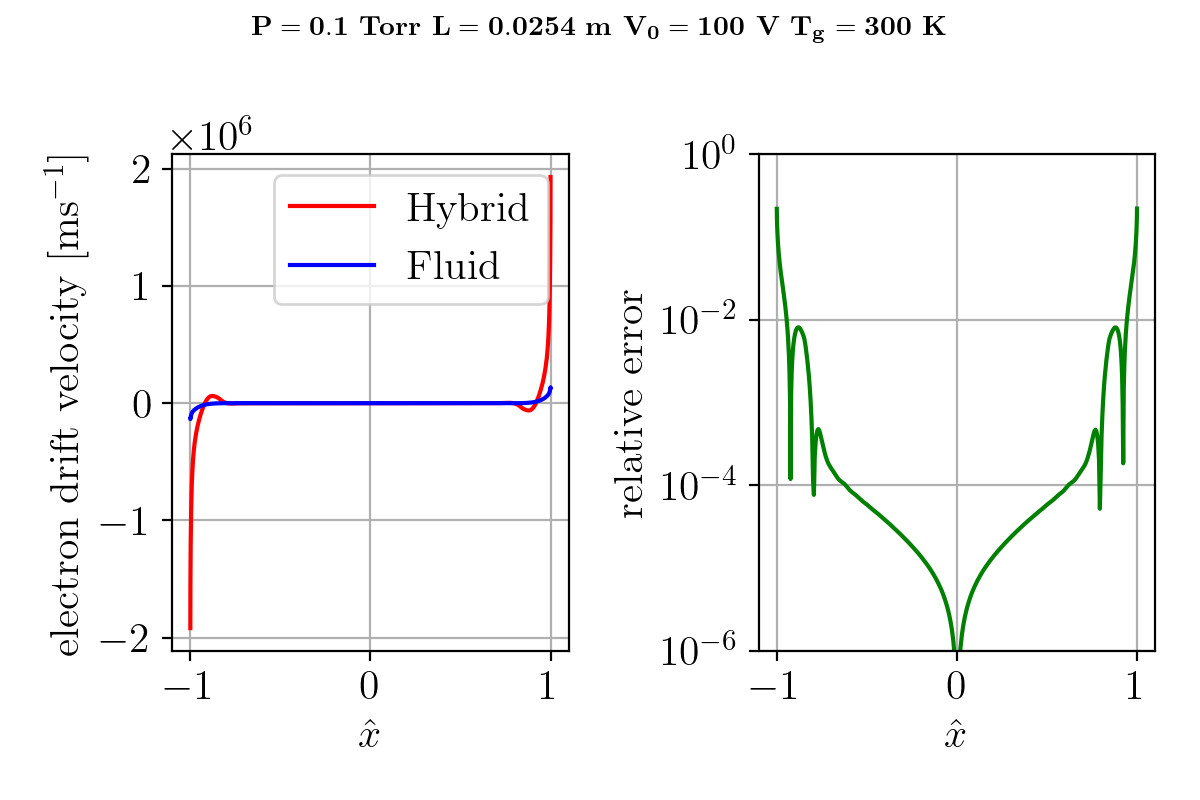} 
	\includegraphics[width=0.46\textwidth]{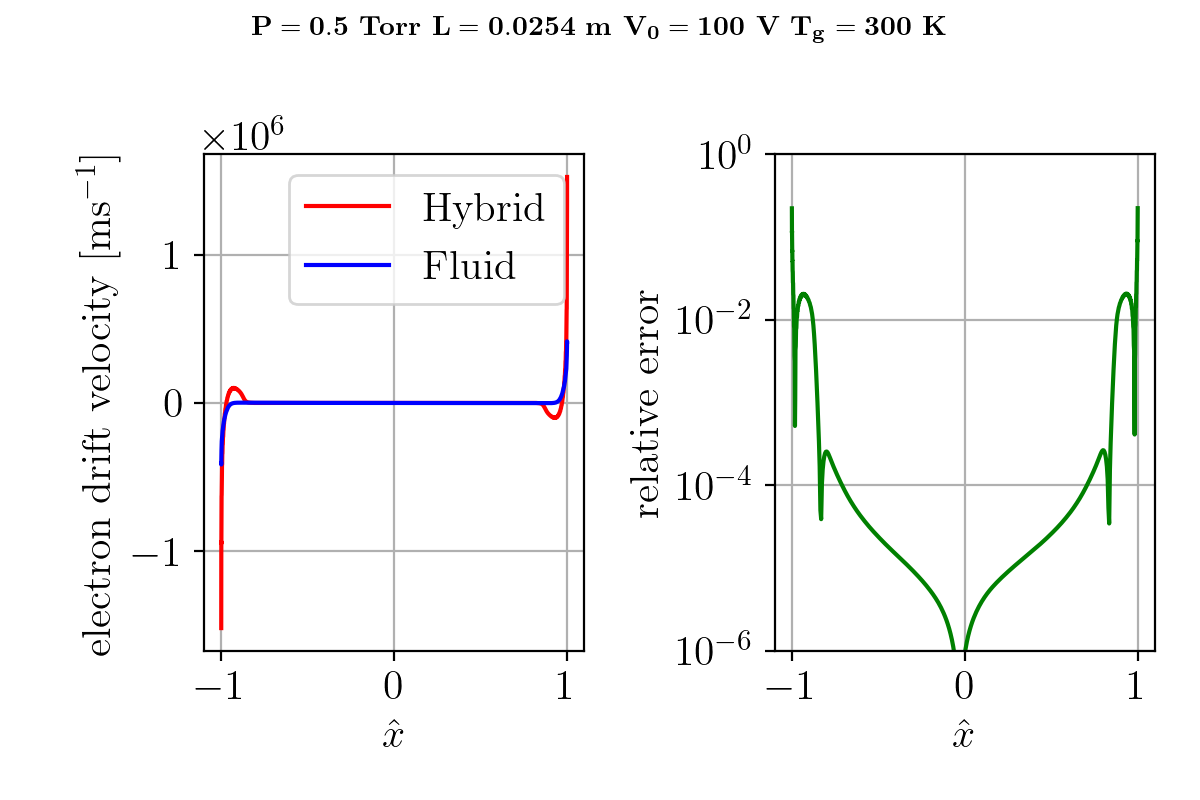}
	\includegraphics[width=0.46\textwidth]{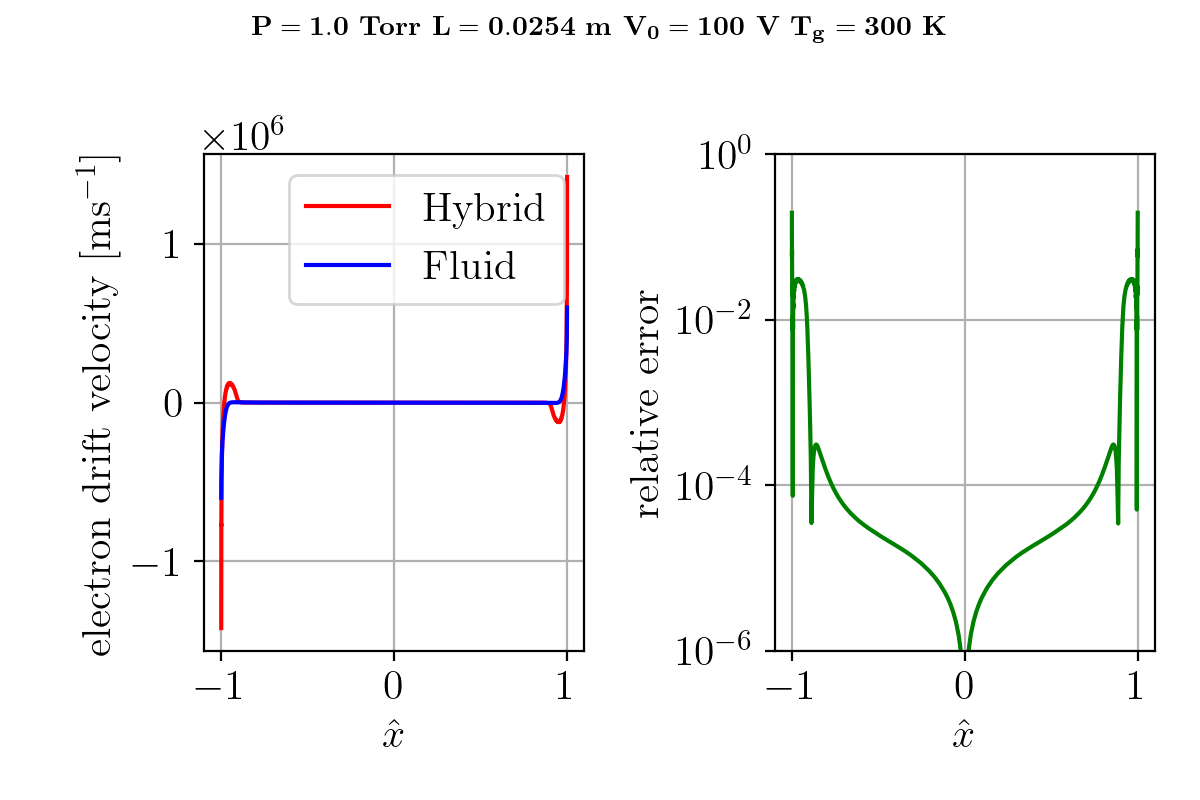}   
	\includegraphics[width=0.46\textwidth]{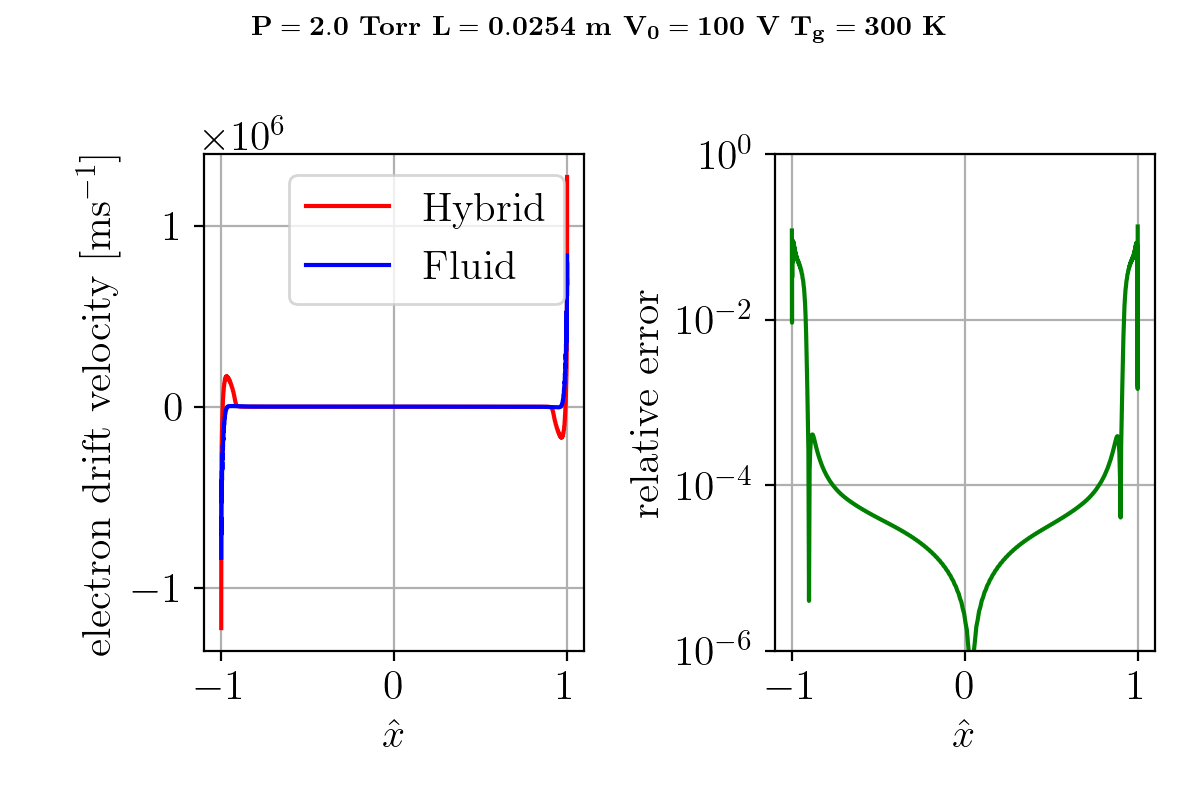}   
	
	\caption{Comparison of the cycle-averaged electron drift velocity computed based on the hybrid and the fluid approximation modeling approaches. The second column for each case shows the relative error between the two approaches taking the hybrid model as the true solution with respect to its $l_2$ norm. \label{fig:drift_velocity_hybrid_fluid}}
\end{figure}

\textbf{The drift-diffusion approximation}: 
The fluid model uses the drift-diffusion approximation to derive a closed expression for the electron flux term $\vect{J}_e$. The drift-diffusion approximation assumes that the EDFs are isotropic and electron drift velocity is negligible compared to the collisional momentum transfer frequency. \Cref{fig:drift_velocity_hybrid_fluid} shows the cycle-averaged electron drift velocity computed by the hybrid and the fluid modeling approaches. At the sheath region, we observe higher relative errors, greater than 1\%, compared to the center. Due to the large electric field effects, the EDFs at the sheath regions depict higher anisotropic effects compared to the discharge center (see \Cref{fig:eedf_hybrid}). Therefore, the drift-diffusion approximation becomes less valid at the sheath regions and has larger errors compared to the discharge center.   

%The presented fluid model uses the drift-diffusion approximation to derive a closed expression for the electron flux term $\vect{J}_e$. The drift-diffusion approximation assumes that the EDFs are isotropic and electron drift velocity is negligible compared to the collisional momentum transfer frequency. \Cref{fig:drift_velocity_hybrid_fluid} shows the electron drift velocity magnitude $\norm{\vect{u}_e}_2$ computed by the hybrid and the fluid modeling approaches. At the sheath region, we observe higher relative errors (i.e., greater than 1\%) compared to the center. Due to the large electric field effects, the EDFs at the sheath regions depict higher anisotropic effects compared to the discharge center (see \Cref{fig:eedf_hybrid}). Therefore, the drift-diffusion approximation becomes less valid at the sheath regions and has larger errors compared to the discharge center.   

\textbf{Boundary conditions}: For the hybrid approximation, we use zero incoming flux boundary conditions for the electrons. This will cause non-Maxwellian EDFs at the discharge walls. In contrast, the fluid approximation assumes Maxwellian EDF at discharge walls. Specifically, for the fluid model, the electron boundary flux term is derived based on the Maxwellian EDF assumption, and it is non-linearly coupled to the electron energy equation. The above discrepancy in the boundary EDF between the models also contributes to the differences we observe in the steady-state profiles.

%\begin{figure}[!tbhp]
%	\centering
%	\includegraphics[width=\textwidth]{fig/0.1Torr300K_100V_Ar_3sp2r_cycle.png} \\
%	\includegraphics[width=\textwidth]{fig/0.5Torr300K_100V_Ar_3sp2r_cycle.png} \\
%	\includegraphics[width=\textwidth]{fig/1Torr300K_100V_Ar_3sp2r_cycle.png}   \\
%	\includegraphics[width=\textwidth]{fig/2Torr300K_100V_Ar_3sp2r_cycle.png}   
%%	\caption{Comparison of the electron drift velocity magnitude ($||\vect{u}_e||_2$) computed based on the hybrid and the fluid approximation modeling approaches. The third column shows the relative error between the two approaches taking the hybrid model as the true solution. \label{fig:drift_velocity_hybrid_fluid}}
%\end{figure}

%\begin{figure}[!tbhp]
%	\includegraphics[width=0.99\textwidth]{fig/0.1Torr300K_100V_Ar_3sp2r_cycle.png} \\
%	\includegraphics[width=0.99\textwidth]{fig/0.5Torr300K_100V_Ar_3sp2r_cycle.png} \\
%	\includegraphics[width=0.99\textwidth]{fig/1Torr300K_100V_Ar_3sp2r_cycle.png} \\
%	\includegraphics[width=0.99\textwidth]{fig/2Torr300K_100V_Ar_3sp2r_cycle.png} \\
%\end{figure}

%
\section{Conclusions} \label{sec:conclusion}
We presented the \bte~solver for the one-dimensional electron BTE for LTPs. \bte~u-ses a Eulerian approach with a Chebyshev-collocation in the position space. In the velocity space, we use a Galerkin discretization in the radial direction and a mixed spherical harmonics with discrete ordinate method for the angular directions. We compare two models for electron transport, fluid and BTE. In the fluid approximation, we rely on the continuity equations for species transport. For the hybrid approximation, we use continuity equations for ions while using the BTE for electron transport. In the fluid approximation, we use non-Maxwellian treatment of EDFs with tabulated kinetic coefficients. As expected, at lower pressures, we observe significant differences between the time-periodic profiles, computed based on the fluid and hybrid approximations. The above indicates that the kinetic treatment of electrons is crucial for accurate RF-GDPs simulation, and the widely-used tabulated interpolation of kinetic coefficients is inadequate to capture the spatial coupling effects on EDFs.

%\begin{itemize}
%	\item What we did in this paper ? 
%	\item The kinetic treatment of electrons is important in LTPs. Alternative modeling approaches with tabulated kinetic coefficients makes significant errors. 
%	\item 
%\end{itemize}
%
\section{Future work}
We plan to explore electron transport differences with more heavy species, especially meta-stables and excited species. 
%Also, we plan to develop a 2D $\vect{x}$-space BTE solver and study two-dimensional fluid and BTE effects on electron transport in LTPs. 
Both fluid and hybrid approaches require long time horizons with thousands of RF cycles to achieve time-periodic solutions. The above can be eliminated by a reduce-order model that can generate approximate time-periodic solutions, and the above can be used as an initial condition for the high-fidelity model. Another possible extension would be to explore spatially adaptive velocity space discretization. This can benefit greatly, especially moving towards solving the 3D3V BTE.

\bibliographystyle{siamplain}
\bibliography{bte}

\begin{thebibliography}{10}

\bibitem{barnes1988staggered}
{\sc M.~S. Barnes, T.~J. Cotler, and M.~E. Elta}, {\em A staggered-mesh
  finite-difference numerical method for solving the transport equations in low
  pressure rf glow discharges}, Journal of Computational physics, 77 (1988),
  pp.~53--72.

\bibitem{bartel2003modelling}
{\sc T.~J. Bartel}, {\em Modelling neutral \& plasma chemistry with dsmc}, in
  AIP Conference Proceedings, vol.~663, American Institute of Physics, 2003,
  pp.~849--856.

\bibitem{boeuf1987numerical}
{\sc J.-P. Boeuf}, {\em Numerical model of rf glow discharges}, Physical review
  A, 36 (1987), p.~2782.

\bibitem{bogaerts1999hybrid}
{\sc A.~Bogaerts, R.~Gijbels, and W.~WimGoedheer}, {\em Hybrid modeling of a
  capacitively coupled radio frequency glow discharge in argon: Combined monte
  carlo and fluid model}, Japanese journal of applied physics, 38 (1999),
  p.~4404.

\bibitem{ciarlet2002finite}
{\sc P.~G. Ciarlet}, {\em The finite element method for elliptic problems},
  SIAM, 2002.

\bibitem{fernando0DBTE}
{\sc M.~Fernando, D.~Bochkov, J.~Almgren-Bell, T.~Oliver, R.~Moser,
  P.~Varghese, L.~Raja, and G.~Biros}, {\em A fast solver for the spatially
  homogeneous electron boltzmann equation}, 2024,
  \url{https://arxiv.org/abs/2409.00207},
  \url{https://arxiv.org/abs/2409.00207}.

\bibitem{frezzotti2011solving}
{\sc A.~Frezzotti, G.~P. Ghiroldi, and L.~Gibelli}, {\em Solving the boltzmann
  equation on gpus}, Computer Physics Communications, 182 (2011),
  pp.~2445--2453.

\bibitem{hagelaar2015coulomb}
{\sc G.~Hagelaar}, {\em Coulomb collisions in the boltzmann equation for
  electrons in low-temperature gas discharge plasmas}, Plasma Sources Science
  and Technology, 25 (2015), p.~015015.

\bibitem{hagelaar2005solving}
{\sc G.~Hagelaar and L.~C. Pitchford}, {\em Solving the boltzmann equation to
  obtain electron transport coefficients and rate coefficients for fluid
  models}, Plasma sources science and technology, 14 (2005), p.~722.

\bibitem{hill1986introduction}
{\sc T.~L. Hill}, {\em An introduction to statistical thermodynamics}, Courier
  Corporation, 1986.

\bibitem{kothnur2007simulation}
{\sc P.~Kothnur and L.~Raja}, {\em Simulation of direct-current microdischarges
  for application in electro-thermal class of small satellite propulsion
  devices}, Contributions to Plasma Physics, 47 (2007), pp.~9--18.

\bibitem{levko2021vizgrain}
{\sc D.~Levko, R.~R. Upadhyay, A.~Karpatne, D.~Breden, K.~Suzuki, V.~Topalian,
  C.~Shukla, and L.~L. Raja}, {\em {VizGrain}: a new computational tool for
  particle simulations of reactive plasma discharges and rarefied flow
  physics}, Plasma Sources Science and Technology, 30 (2021), p.~055012.

\bibitem{liu2014numerical}
{\sc Q.~Liu, Y.~Liu, T.~Samir, and Z.~Ma}, {\em Numerical study of effect of
  secondary electron emission on discharge characteristics in low pressure
  capacitive rf argon discharge}, Physics of Plasmas, 21 (2014).

\bibitem{loffhagen2009advances}
{\sc D.~Loffhagen and F.~Sigeneger}, {\em Advances in boltzmann equation based
  modelling of discharge plasmas}, Plasma Sources Science and Technology, 18
  (2009), p.~034006.

\bibitem{lymberopoulos1993fluid}
{\sc D.~P. Lymberopoulos and D.~J. Economou}, {\em Fluid simulations of glow
  discharges: Effect of metastable atoms in argon}, Journal of applied physics,
  73 (1993), pp.~3668--3679.

\bibitem{lymberopoulos1994stochastic}
{\sc D.~P. Lymberopoulos and J.~D. Schieber}, {\em Stochastic dynamic
  simulation of the boltzmann equation for electron swarms in glow discharges},
  Physical Review E, 50 (1994), p.~4911.

\bibitem{oblapenko2020velocity}
{\sc G.~Oblapenko, D.~Goldstein, P.~Varghese, and C.~Moore}, {\em A velocity
  space hybridization-based boltzmann equation solver}, Journal of
  Computational Physics, 408 (2020), p.~109302.

\bibitem{panneer2015computational}
{\sc P.~K. Panneer~Chelvam and L.~L. Raja}, {\em Computational modeling of the
  effect of external electron injection into a direct-current microdischarge},
  Journal of Applied Physics, 118 (2015).

\bibitem{pitchford2017lxcat}
{\sc L.~C. Pitchford, L.~L. Alves, K.~Bartschat, S.~F. Biagi, M.-C. Bordage,
  I.~Bray, C.~E. Brion, M.~J. Brunger, L.~Campbell, A.~Chachereau, et~al.},
  {\em Lxcat: An open-access, web-based platform for data needed for modeling
  low temperature plasmas}, Plasma Processes and Polymers, 14 (2017),
  p.~1600098.

\bibitem{satake1997two}
{\sc K.~S.~K. Satake, T.~M.~T. Monaka, O.~U.~O. Ukai, Y.~T.~Y. Takeuchi, and
  M.~M.~M. Murata}, {\em Two-dimensional nonequilibrium plasma modeling based
  on the particle-boltzmann hybrid model for rf glow discharges}, Japanese
  journal of applied physics, 36 (1997), p.~4789.

\bibitem{scanlon2010open}
{\sc T.~Scanlon, E.~Roohi, C.~White, M.~Darbandi, and J.~Reese}, {\em An open
  source, parallel dsmc code for rarefied gas flows in arbitrary geometries},
  Computers \& Fluids, 39 (2010), pp.~2078--2089.

\bibitem{surendra1991particle}
{\sc M.~Surendra and D.~B. Graves}, {\em Particle simulations of
  radio-frequency glow discharges}, IEEE transactions on plasma science, 19
  (1991), pp.~144--157.

\bibitem{tsendin1995electron}
{\sc L.~Tsendin}, {\em Electron kinetics in non-uniform glow discharge
  plasmas}, Plasma Sources Science and Technology, 4 (1995), p.~200.

\bibitem{wilcoxson1996simulation}
{\sc M.~H. Wilcoxson and V.~I. Manousiouthakis}, {\em Simulation of a
  three-moment fluid model of a two-dimensional radio frequency discharge},
  Chemical engineering science, 51 (1996), pp.~1089--1106.

\bibitem{young1993comparative}
{\sc F.~F. Young et~al.}, {\em A comparative study between non-equilibrium and
  equilibrium models of rf glow discharges}, Journal of Physics D: Applied
  Physics, 26 (1993), p.~782.

\bibitem{young1993two}
{\sc F.~F. Young and C.-H. Wu}, {\em Two-dimensional, self-consistent,
  three-moment simulation of rf glow discharge}, IEEE transactions on plasma
  science, 21 (1993), pp.~312--321.

\bibitem{yuan20171d}
{\sc C.~Yuan, E.~Bogdanov, S.~Eliseev, and A.~Kudryavtsev}, {\em 1d kinetic
  simulations of a short glow discharge in helium}, Physics of Plasmas, 24
  (2017).

\bibitem{zabelok2015adaptive}
{\sc S.~Zabelok, R.~Arslanbekov, and V.~Kolobov}, {\em Adaptive kinetic-fluid
  solvers for heterogeneous computing architectures}, Journal of Computational
  Physics, 303 (2015), pp.~455--469.

\bibitem{zhao2018numerical}
{\sc L.-L. Zhao, Y.~Liu, and T.~Samir}, {\em Numerical study on discharge
  characteristics influenced by secondary electron emission in capacitive rf
  argon glow discharges by fluid modeling}, Chinese Physics B, 27 (2018),
  p.~025201.

\end{thebibliography}
\end{document}